\documentclass[sigconf,screen,twocolumn]{acmart}

\renewcommand\footnotetextcopyrightpermission[1]{} 
\usepackage{comment}
\usepackage{graphicx}
\usepackage{subfig}
\usepackage{multirow}
\usepackage{float}
\usepackage{xcolor}
\usepackage{enumitem}
\usepackage{url}
\usepackage{romannum}
\usepackage{array}
\usepackage{makecell}
\usepackage{amsmath,bm}
\usepackage[normalem]{ulem}

\usepackage{tikz}
\newcommand*\circled[1]{\tikz[baseline=(char.base)]{
            \node[shape=circle,draw,inner sep=0.6pt] (char) {#1};}}

\AtBeginDocument{%
   \providecommand\BibTeX{{%
     \normalfont B\kern-0.5em{\scshape i\kern-0.25em b}\kern-0.8em\TeX}}}
\setcopyright{none}            
\begin{document}

\title{JOSS: Joint Exploration of CPU-Memory DVFS and Task Scheduling for Energy Efficiency} 




\author{Jing Chen}
\affiliation{%
  \institution{Chalmers University of Technology}
  \city{Gothenburg}
  \country{Sweden}}
\email{chjing@chalmers.se}

\author{Madhavan Manivannan}
\affiliation{%
  \institution{Chalmers University of Technology}
  \city{Gothenburg}
  \country{Sweden}}
\email{madhavan@chalmers.se}

\author{Bhavishya Goel}
\affiliation{%
  \institution{Chalmers University of Technology}
  \city{Gothenburg}
  \country{Sweden}}
\email{goelb@chalmers.se}

\author{Miquel Pericàs}
\affiliation{%
  \institution{Chalmers University of Technology}
  \city{Gothenburg}
  \country{Sweden}}
\email{miquelp@chalmers.se}

\begin{abstract}

Energy-efficient execution of task-based parallel applications is crucial as tasking is a widely supported feature in many parallel programming libraries and runtimes. 
Currently, state-of-the-art proposals primarily rely on leveraging core asymmetry and CPU DVFS.
Additionally, these proposals mostly use heuristics and lack the ability to explore the trade-offs between energy usage and performance.
However, our findings demonstrate that focusing solely on CPU energy consumption for energy-efficient scheduling while neglecting memory energy consumption leaves room for further energy savings. 
We propose JOSS, a runtime scheduling framework that leverages both CPU DVFS and memory DVFS in conjunction with core asymmetry and task characteristics to enable energy-efficient execution of task-based applications. JOSS also enables the exploration of energy and performance trade-offs by supporting user-defined performance constraints.
JOSS uses a set of models to predict task execution time, CPU and memory power consumption, and then selects the configuration for the tunable knobs to achieve the desired energy performance trade-off. 
Our evaluation shows that JOSS achieves 21.2\% energy reduction, on average, compared to the state-of-the-art. 
Moreover, we demonstrate that even in the absence of a memory DVFS knob,
taking energy consumption of both CPU and memory into account achieves better energy savings compared to only accounting for CPU energy.
Furthermore, \textcolor{black}{JOSS is able to adapt scheduling to reduce energy consumption while satisfying the desired performance constraints.}

\end{abstract}
\maketitle
\section{Introduction} \label{introduction}
Energy efficiency has emerged as a crucial design constraint in various parallel computing systems ranging from battery-powered mobile devices to high performance computers.
Modern chip multi - processors (CMPs) incorporate a variety of hardware features to improve energy efficiency.
The integration of 
multiple core types on a single die, referred to as static asymmetry, in big.LITTLE architectures~\cite{AppleA16, JetsonTX2Module, IntelAlderLake, odroid-xu4, arm-big-little}, offers the possibility to execute applications on different cores with varied performance and power characteristics.
Modern CMPs also support dynamic voltage and frequency scaling (DVFS), 
referred to as dynamic asymmetry, which manages system power consumption and enables exploration of performance and power consumption trade-offs when executing applications. 
In addition, 
cores of the same type are typically grouped into clusters to reduce the design cost and complexity associated with enabling per-core DVFS~\cite{PerCorevsClustered}.
In such core-clustered designs, the cores in a single cluster operate at the same DVFS setting~\cite{IntelAlderLake, odroid-xu4, JetsonTX2Module}. 

There has been extensive research on leveraging static and/or dynamic asymmetry, 
as knobs to reduce CPU energy consumption since it is typically the largest contributor to the total energy consumption in a system~\cite{Modeling/Tradeoff/SC2018, EnergyMinimization/OpenMP/DVFS/DCT/PARMA-DITAM2015, HarisRibic-AEQUITAS, Christopher-ISCA2016-Asymmetric/PercoreDVFS/Parallelism, EmilioCastillo-IPDPS2016-CATA/Criticality/EDP, HMP/DVFS/Criticality/Ratio/Prediction/TPDS2020,DCT/DVFS/Task-base/ICPP21, Tradeoff/hetero/model/Energy2020}.
In order to meet the memory demand for emerging many-core architectures, main memory bandwidth and capacities have also been steadily increasing (albeit at a slower rate)\cite{MemSpeed}.
This has led to the memory system also becoming a major contributor to the total energy consumption~\cite{MemoryEnergyIncrease}.
Several works have highlighted the importance and the benefit of leveraging memory DVFS, in conjunction with static and/or dynamic asymmetry offered by the CPUs, since it opens up many opportunities for establishing trade-offs between performance and energy consumption~\cite{MemScale/ASPLOS2011, Coscale/MICRO2012, JointDVFS/Supercomputing16,MemDVFS/ICAC2011}. These proposals however focus on the potential of leveraging these knobs in the context of single-threaded and multi-programmed workloads (comprising several single-threaded applications).




Task-based parallel programming models are supported by many production parallel programming libraries and runtimes~\cite{openmp50-api, StarPU, frigo-cilk5, tbb_scheduler}, such as OpenMP, since it eases the process of expressing parallelism inherent in applications.
In this model, an application is expressed as a directed acyclic graph (DAG), where the tasks (vertex) and their dependencies (edges) are generated dynamically during execution. 
Tasks can be of different types and exhibit diverse attributes (e.g.~OPs/byte ratio)~\cite{BOTS/ICPP2009}.
Additionally, moldable execution enabled by intra-task parallelism (i.e.~using multiple cores for running a single task) has been shown to improve performance
and lower the impact of system idle energy~\cite{Jing_ERASE, STEER/SBACPAD2022}.



Energy-efficient execution of a task DAG relies on runtime schedulers to map each task in the DAG to hardware resources (i.e.~choosing the appropriate core type and number of cores for the task) and throttle the available DVFS knobs simultaneously. 
Several recent works have targeted energy-efficient runtime scheduling techniques for task-based parallel applications, which can be broadly grouped into three categories~\cite{Jing_ERASE,TaskAwarenessSchedule/ICA3PP2017, DCT/DVFS/Task-base/ICPP21,HarisRibic-AEQUITAS, EmilioCastillo-IPDPS2016-CATA/Criticality/EDP, HMP/DVFS/Criticality/Ratio/Prediction/TPDS2020,SchedDec/DVFS/CPE2019, Christopher-ISCA2016-Asymmetric/PercoreDVFS/Parallelism, Tradeoff/hetero/model/Energy2020, STEER/SBACPAD2022}.
The first category primarily exploits task characteristics (e.g.~task OPs/byte, task size, task criticality) in conjunction with static asymmetry without leveraging DVFS~\cite{Jing_ERASE,TaskAwarenessSchedule/ICA3PP2017}. 
The second category employs task characteristics in conjunction with dynamic asymmetry, while being restricted to symmetric architectures~\cite{HarisRibic-AEQUITAS, HarisRibic-WorkStealing/PercoreDVFS/Tempo, DCT/DVFS/Task-base/ICPP21}. 
The third category employs task characteristics in conjunction with static asymmetry and a subset of dynamic asymmetry knobs (CPU DVFS) for scheduling~\cite{EmilioCastillo-IPDPS2016-CATA/Criticality/EDP, Christopher-ISCA2016-Asymmetric/PercoreDVFS/Parallelism, HMP/DVFS/Criticality/Ratio/Prediction/TPDS2020, DCT/DVFS/Task-base/ICPP21, Tradeoff/hetero/model/Energy2020, STEER/SBACPAD2022}. 

Unfortunately, existing works do not leverage static and dynamic asymmetry, especially memory DVFS, 
in conjunction with task characteristics 
for scheduling.
Our analysis in Section~\ref{motivation} shows that 
utilizing all the knobs, i.e.~core type ($\mathbf{T_{C}}$), number of cores ($\mathbf{N_{C}}$), core frequency ($\bm{f_{C}}$) and memory frequency ($\bm{f_{M}}$) in conjunction, provides greater opportunities for reducing energy consumption over using a subset of the knobs.
We also show that, even in the absence of a memory DVFS knob, considering CPU energy consumption alone, as done in prior works, 
for scheduling tasks without taking memory energy consumption into account leaves a lot of scope for improvement. 
Furthermore, prior art is designed for a specific target and mostly uses a set of heuristics without having the ability to explore trade-offs between performance and energy consumption. 

The goal is to develop a runtime scheduling framework that leverages the aforementioned 
knobs and provides the ability to target various trade-offs between performance and energy consumption. 
To achieve this goal, several challenges need to be addressed. Firstly,  the effects of tuning available 
knobs, individually and in conjunction, on performance and energy consumption needs to be understood. Secondly, the interplay between
task scheduling decisions and the exploration of various trade-offs between energy and performance needs to be investigated. 
Thirdly, there is a need to accommodate applications/tasks with diverse characteristics 
while also ensuring low runtime overhead. 

We propose \textbf{JOSS} (\underline{\textbf{JO}}int  \underline{\textbf{S}}cheduling and \underline{\textbf{S}}caling), a runtime scheduling framework that can target  
various energy performance trade-offs through leveraging the aforementioned knobs. 
Overall, JOSS achieves the goals set out by optimizing the execution of 
each task in the DAG.
For instance, JOSS reduces the total energy consumption of a task-based application by running each task with the lowest energy possible. 
\textcolor{black}{The operation of the framework can be summarized as follows. 
First, JOSS utilizes multivariate polynomial regression based models for predicting} the execution time and power consumption of CPU and memory subsystem when running a task on different configurations spanning the four knobs <$\mathrm{T_{C}}$, $\mathrm{N_{C}}$, $f_{C}$, $f_M$>.
Second, the JOSS task scheduler combines the model predictions in conjunction with instantaneous task concurrency during runtime to estimate the energy consumption of a task and make the scheduling decisions accordingly for various trade-off goals.  
To prune the large search space formed by the four knobs and reduce the overhead during runtime, JOSS uses a steepest descent approach to determine the configuration that satisfies the desired trade-off in a few steps. 

We evaluate JOSS under two scenarios targeting different energy performance trade-offs: \circled{1} reducing the total energy consumption and \circled{2} reducing energy with user-specified performance speedup with respect to \circled{1}.
The results for scenario \circled{1} indicate that JOSS achieves 21.2\% energy reduction on average compared to the state-of-the-art~\cite{STEER/SBACPAD2022} on an NVIDIA Jetson TX2 platform. 
Even in the absence of a memory DVFS knob, JOSS can still provide an additional 5.2\% energy reduction than the state-of-the-art. 
For scenario \circled{2}, we show that JOSS is able to adapt scheduling to reduce energy consumption while satisfying the desired performance constraints. 

In summary, this paper makes the following contributions:
\begin{itemize}[topsep=1pt,leftmargin=*]
    \item We demonstrate that (1) leveraging static asymmetry and dynamic asymmetry, i.e.~core type, number of cores, core frequency and memory frequency, in conjunction with task characteristics, enables significant reduction in energy consumption; 
    (2) even in the absence of a memory DVFS knob, taking total energy consumption (including CPU and memory) into account for configuration selection results in lower energy consumption compared to only considering CPU energy. 
    \item We propose the JOSS runtime scheduling framework for task-based parallel applications on multicore architectures, which provides the ability to explore various energy performance trade-offs. 
    \item 
    We build a set of models using multivariate polynomial regression capable of accurately predicting the execution time, CPU power, and memory power of each task when tuning the four available knobs, individually and in conjunction. 
\end{itemize}

\vspace{-1mm}
\section{Motivation} \label{motivation}
We motivate JOSS by demonstrating
\textcolor{black}{(1) the importance of taking memory energy consumption into account and its impact on selecting configurations for the different knobs;
(2) the impact of leveraging the four knobs <$\mathrm{T_{C}}, \mathrm{N_{C}}, f_{C}, f_M$> in conjunction on reducing energy consumption;}
(3) the potential for energy performance trade-off exploration (e.g.~reducing energy while satisfying user-specified performance constraints).
We use NVIDIA Jetson TX2 as the experimental platform since it features static asymmetry (i.e.~a dual-core high-performance Denver CPU and a relatively low-performance quad-core A57 CPU) and dynamic asymmetry (CPU and memory DVFS knobs that can be tuned during execution). 

For this experiment, we use two different benchmarks - Matrix Multiplication (MM, compute-intensive) and Memory Copy (MC, memory-intensive), configured with a DAG parallelism ($dop$) of one, where $dop$ represents the potential task concurrency in the task DAG obtained by dividing the total number of tasks by the length of the longest path.
We run both benchmarks with all possible combinations of the knobs, measure CPU and memory power consumption, and execution time and use these results in our analysis.
In these benchmarks, tasks represent the numerical
kernels, which are typically invoked \textcolor{black}{numerous} times wherein the routine(s) executed by different tasks (invocations of the same kernel) are identical. 
Additional details about the platform and the benchmarks are provided in Section~\ref{setup}.

\begin{figure}[!b]
\centering
\vspace{-7mm}
\includegraphics[width=\columnwidth]{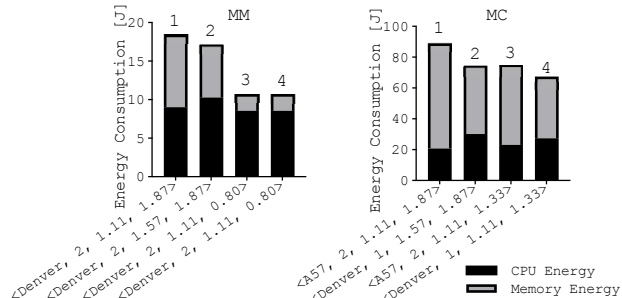}
\vspace{-10mm}
\caption{Total energy consumption under the four scenarios.}
\label{fig:includeMemory}
\end{figure}

\vspace{-1mm}
\subsection{Importance of Including Memory Energy}
\textcolor{black}{To motivate the importance of including memory energy for CPU DVFS,} we compare the total energy consumption (CPU+memory) under 
\textcolor{black}{different scenarios} as shown in Figure~\ref{fig:includeMemory}. 
\textcolor{black}{Scenario 1 ($1^{st}$ bar) represents the state-of-the-art~\cite{STEER/SBACPAD2022}, where  
the configuration that consumes the least CPU energy is identified while tuning the three knobs <$\mathrm{T_{C}}, \mathrm{N_{C}}, f_{C}$>} and fixing $f_{M}$ at the highest frequency 1.87GHz.
In scenario 2 ($2^{nd}$ bar), we identify the configuration that consumes the least total energy while tuning <$\mathrm{T_{C}}, \mathrm{N_{C}}, f_{C}$> and fixing $f_{M}$ at 1.87GHz.
Note that for both scenarios, we assume that memory DVFS knob is unavailable and that the memory always operates at the highest frequency. 
From Figure~\ref{fig:includeMemory} we can observe that the configuration identified (see x-axis labels) when only considering CPU energy consumption is sub-optimal 
for both MM and MC. 
However, taking memory energy into account leads to a different configuration that further reduces the total energy, while still being restricted to three knobs.
For instance, in the case of MC the best configuration changes from <A57, 2, 1.11> to <Denver, 1, 1.57> and leads to 16\% energy reduction.

\vspace{-1mm}
\subsection{\textcolor{black}{Leveraging Knobs in Conjunction}}
\textcolor{black}{To motivate the importance of leveraging the four knobs in conjunction, we compare the total energy consumption under scenarios 3 and 4 in Figure~\ref{fig:includeMemory}}. 
Here we assume the memory DVFS knob to be tunable.
Scenario 3 ($3^{rd}$ bar) \textcolor{black}{enhances the state-of-the-art~\cite{STEER/SBACPAD2022} with support for orthogonal frequency scaling} where CPU and memory frequencies are throttled independently. 
This involves determining the best configuration of <$\mathrm{T_{C}}, \mathrm{N_{C}}, f_{C}$>, as in scenario 1 and then evaluating the total energy consumption when tuning $f_{M}$ while the other three knobs remain fixed. 
Scenario 4 ($4^{th}$ bar) represents 
\textcolor{black}{leveraging the four knobs in conjunction (the approach adopted by JOSS)}
where energy consumption for the entire configuration space is searched to determine the configuration that consumes the least total energy. 
The results show that in the case of MC 
\textcolor{black}{scenario 4}
ends up selecting a different configuration and leads to 10\% energy savings compared to scenario 3. 
For MM
\textcolor{black}{scenario 4}
does not provide any additional benefit since there is no change in the configuration.

In summary, we conclude that \textit{(i) even in the absence of a memory DVFS knob, only considering CPU energy results in sub-optimal configuration selection thereby emphasizing the importance of also taking memory energy consumption into account; 
(ii) in comparison to the orthogonal CPU DVFS and memory DVFS throttling, 
\textcolor{black}{leveraging the four knobs in conjunction}
can lead to more energy savings.
}



\begin{figure}[!t]
\centering
\vspace{-4mm}
\includegraphics[width=\columnwidth]{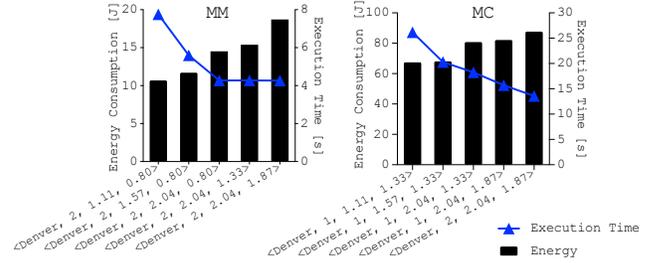}
\vspace{-9mm}
\caption{Energy performance trade-offs exploration.}
\vspace{-4mm}
\label{fig:EPTO}
\end{figure}

\vspace{-1mm}
\subsection{Exploring Energy Performance Trade-offs} 
It is crucial that the scheduler aims to reduce energy consumption while still maintaining a good level of performance. 
Figure~\ref{fig:EPTO} shows the potential trade-offs between energy consumption and performance when tuning the available knobs.  
Note that the first bar represents the configuration that consumes the least total energy consumption which we will use as the baseline for the rest of this discussion.
We can observe that
tuning core frequency from 1.11 to 1.57 for MM and MC provides 1.4$\times$ and 1.3$\times$ performance speedup while increasing energy consumption by 10\% and 1\%, respectively.
The maximum speedup for MM is 1.8$\times$ which comes at the cost of 36\% increase in energy consumption. 
 MC can achieve a maximum speedup of 
1.9$\times$ at the cost of 30\% increase in energy consumption. 
\textit{Building a runtime scheduling framework that can flexibly explore energy performance trade-offs will enable the user to customize the scheduler to their specific requirements.}

\vspace{-1mm}
\section{JOSS Runtime Framework Overview}
\begin{figure}[!t]
\centering
\includegraphics[width=0.9\columnwidth]{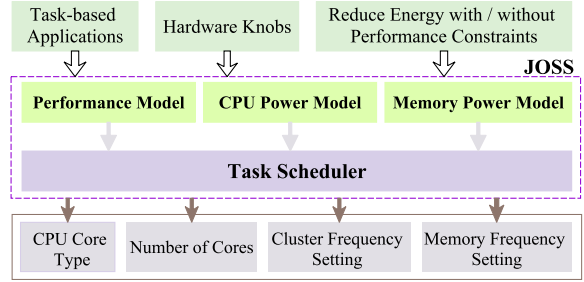}
\vspace{-3mm}
\caption{High-level overview of JOSS.
}
\vspace{-6mm}
\label{fig:overview}
\end{figure}

Figure~\ref{fig:overview} provides a high-level overview of the JOSS runtime framework. The framework takes
the application, details about the architectural knobs (number of clusters, number of cores in each cluster, supported frequencies) and performance constraints
(that can be specified by either user or system software) as inputs.


To enable energy performance trade-off exploration, JOSS leverages four knobs <$\mathrm{T_{C}}, \mathrm{N_{C}}, f_{C}, f_{M}$> whose effects on energy and performance need to be understood when tuned individually and in combination.
We therefore develop performance and power models 
to understand the effects and guide configuration selection. JOSS specifically comprises 
a performance model, a CPU power model and a memory power model to provide the scheduler with the prediction of task execution time and power consumed in CPU and memory domains when executing tasks with different configurations. 
Energy and performance estimates are used by the task scheduler to guide  configuration selection and achieve the desired trade-off goals.
The scheduler maps tasks to selected CPU cores and sends the frequency throttling requests to the CPU DVFS controller and the memory DVFS controller. 
JOSS targets total energy reduction if a performance constraint is not specified.

In Section~\ref{models}, we first introduce the development of the three prediction models in JOSS.
Section~\ref{design} will then describe the task scheduler and how these models are utilized 
to 
enable energy performance trade-off exploration.



\vspace{-1mm}
\section{Models} \label{models}
\begin{figure*}[!t]
\centering
\vspace{-3mm}
\includegraphics[width=0.9\textwidth]{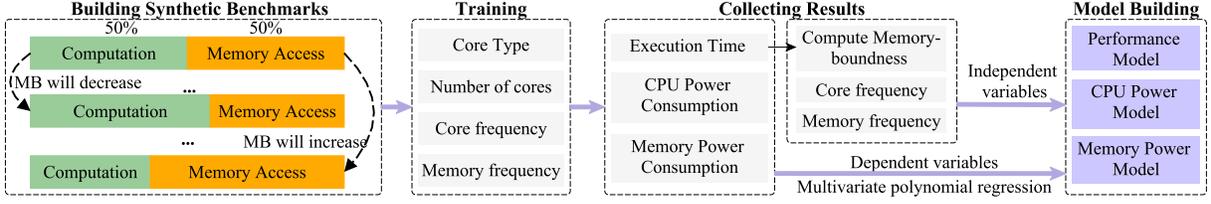}
\vspace{-3mm}
\caption{Overview of the model building process. }
\vspace{-4mm}
\label{fig:modelbuilding}
\end{figure*}


\textbf{Challenges:} JOSS utilizes models to predict performance and power consumption of tasks and understand the energy and performance effects from tuning the four knobs <$\mathrm{T_{C}}, \mathrm{N_{C}}$, $f_{C}, f_{M}$>.
Two challenges need to be addressed in this context. 

First, creating a model to predict the effect of four tunable knobs for varying task characteristics (i.e.~five parameters in total) is complicated and expensive, especially as the number of configurations increases. To tackle this issue, JOSS combines models with runtime samples in a hybrid approach. 
We sample the two knobs <$\mathrm{T_{C}}, \mathrm{N_{C}}$>
during runtime (details in Section~\ref{SamplePrediction}) and use models to predict the impact of DVFS on performance and power consumption of a task from the other two knobs <$f_{C}, f_{M}$>.

Second, prediction models for execution time and power consumption proposed in prior works
rely on the availability of a number of Performance Monitoring Counters (PMCs)~\cite{goel2016measurement, WASH/regression/PMC/CGO2016,  Regression/powermodel/TCAD2020, Regression/PMC/Perf/Power/MC2016,Asymmetric/Clusterd/MultipleAPP/MRPI, Phases/MPKI/offline/DVFS/Micro2006,SchedDec/DVFS/CPE2019} and/or model the impact of a subset of the four knobs~\cite{Jing_ERASE, STEER/SBACPAD2022}. 
However, the availability of specific PMCs on different architectures limits the adoption of existing models ~\cite{PMC_pitfall_SP2019, PMC_Flaws_NPC2011}.
\textcolor{black}{For instance, Goel et al.~\cite{goel2016measurement} use \texttt{CYCLE:ACTIVITY:STALLS\_L2\_PENDING} on Haswell to estimate the memory intensity of an application, which is unavailable on later Intel microarchitectures.
Even the TX2 platform used in our evaluation does not provide PMCs related to stall cycles.
To address this, the models used in JOSS  
do not rely on any PMCs thereby improving portability across architectures. 
}

\textbf{Overview: }
To enable performance and power predictions, we first characterize the platform by running a set of synthetic benchmarks. 
Since the impact of DVFS on a task depends on its use of computational and memory resources, we generate a set of synthetic benchmarks with varying levels of  utilization of the two components.  
We execute the synthetic benchmarks while tuning all four knobs during the training stage and collect the corresponding execution time and average power consumption values.
Subsequently, we build the performance and power prediction models using multivariate polynomial regression (MPR).
\textcolor{black}{This technique is commonly used to model the implicit nonlinear relationships between variables~\cite{WASH/regression/PMC/CGO2016,  Regression/powermodel/TCAD2020, Regression/PMC/Perf/Power/MC2016}.
}
Figure~\ref{fig:modelbuilding} depicts an overview of the model building process. 

\vspace{-1mm}
\subsection{Synthetic Benchmarking and Profiling} \label{synbench}
The basic structure of the synthetic benchmark includes a computation loop and a memory access loop.
Through controlling the number of iterations in each loop, we can generate different ratios of computations and memory access.
In this work, by keeping the total execution time of synthetic benchmarks constant, we start from 50\% of computation and 50\% of memory access and then increase or decrease the execution time of corresponding
loops by 2.5\% as shown in Figure~\ref{fig:modelbuilding}. 
In total, we generate 41 synthetic benchmarks with different ratios between computation and memory access. 
We profile the platform via executing these synthetic benchmarks at all possible configurations for the four knobs and measure the execution time, CPU and memory power consumption. 

\vspace{-1mm}
\subsection{Performance Model}
The performance model aims to predict the execution time of a task under joint CPU and memory frequency scaling.
In the model, the total execution time is estimated as the sum of computation time and stall time due to memory latency: $\mathrm{Time = Time_{comp} + Time_{stall}}$.
\textcolor{black}{We use \textit{memory-boundness} (MB) to quantify the fraction of time CPU is stalled due to memory latency
~\cite{goel2016measurement, STEER/SBACPAD2022}. 
When scaling core frequency ($f_{C}\rightarrow f_{C}'$),
the computation time will scale linearly. So the computation time $\mathrm{Time_{comp}'}$ at frequency $f_{C}'$ can be calculated as:} 
\vspace{-1mm}
\begin{equation}\label{time_comp}  \footnotesize
    Time_{comp}' = Time \times (1 - MB) \times \frac{f_{C}}{f_{C}'}
    \vspace{-1mm}
\end{equation}

$\mathrm{Time_{stall}'}$ is dependent on core and memory frequency scaling in addition to task characteristics (MB).
Memory frequency scaling directly influences the latency of memory access.
Core frequency scaling, however, has an (indirect) impact on how often a core issues memory access requests. 
We utilize the statistics from running synthetic benchmarks presented in Section~\ref{synbench} to build the performance model using MPR
as shown below:
\vspace{-2mm}
\begin{equation}\label{timep}   \footnotesize
Time_{stall}' = Time \times (\sum\limits_{i=0}^{2} \beta_{i}x_{i} + \sum\limits_{i=0}^{2} \beta_{ii}x_{i}^{2} + \sum\limits_{i=0}^{1}\sum\limits_{k=i+1}^{2} \beta_{ik}x_{i}x_{k} + \varepsilon)
\vspace{-1mm}
\end{equation}
\textcolor{black}{$x_{i}=\{MB, \frac{f_{C}}{f_{C}'}, \frac{f_{M}}{f_{M}'}\},$ 0$\le$i$\le$2.
Here, $\beta_{i}, \beta_{ii}, \beta_{ik}$ are the coefficients of the linear component, the quadratic component and the interaction component, respectively, and $\varepsilon$ is the intercept. 
The execution time at <$f_{C}', f_{M}'$> is $\mathrm{Time' = Time_{comp}' + Time_{stall}'}$.}

Estimating task execution time at different <$f_{C}', f_{M}'$> settings during runtime requires knowing the MB value and a sampled execution time at a reference <$f_{C}, f_{M}$> setting.
To obtain the MB value without using PMCs, we sample the execution times at two different core frequency settings during runtime 
(i.e.~sample $Time$ and $Time'$ at $f_{C}$ and $f_{C}'$) 
under a fixed memory frequency.
We then obtain MB as follows: 
\vspace{-1mm}
\begin{equation}\label{MB} \footnotesize
    MB = (\frac{Time'}{Time} - \frac{f_{C}}{f_{C}'}) / (1 - \frac{f_{C}}{f_{C}'})
    \vspace{-2mm}
\end{equation}


\begin{figure}[!b]
\centering
\vspace{-8mm}
\subfloat[CPU Power]{\includegraphics[width=0.5\columnwidth]{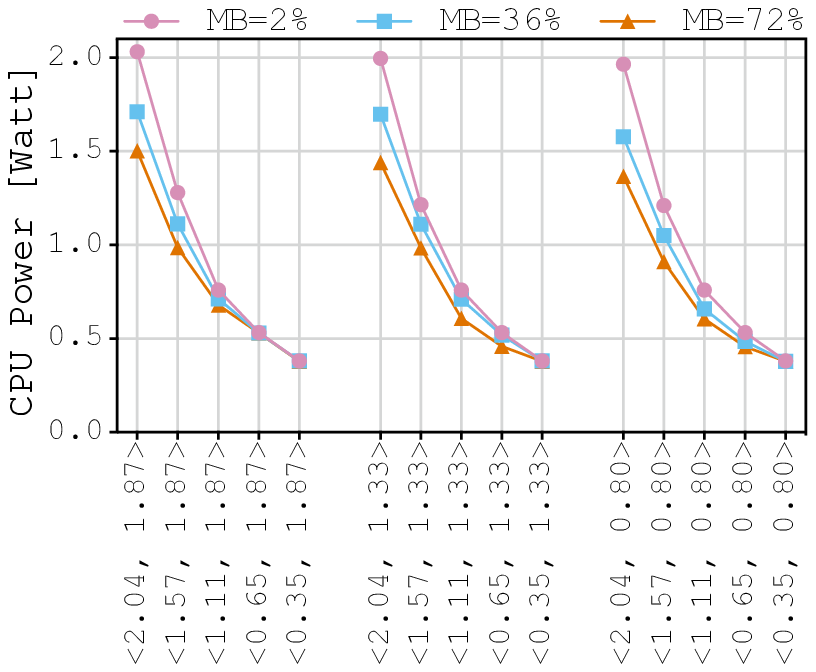}
\label{fig_compow}}
\subfloat[Memory Power]{\includegraphics[width=0.5\columnwidth]{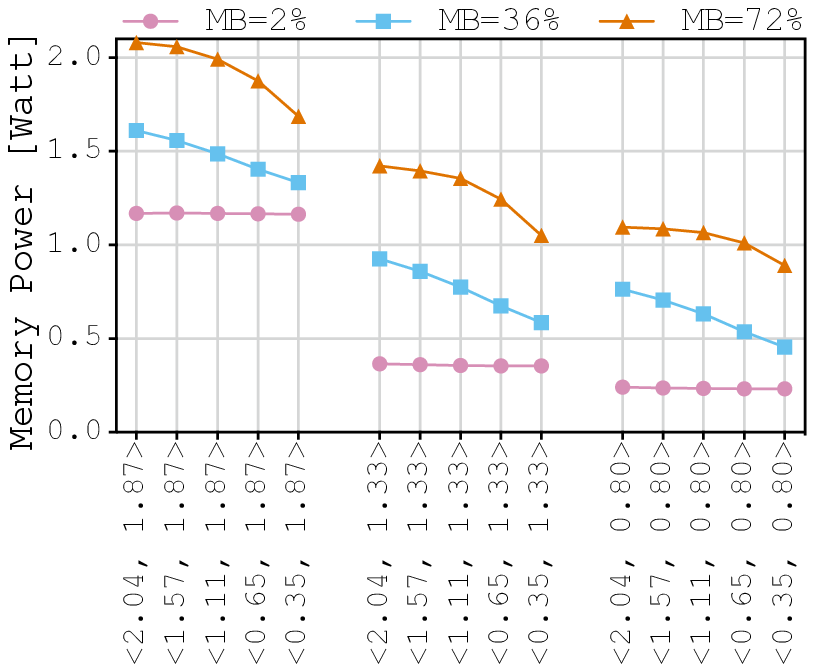}
\label{fig_mempow}}
\vspace{-4mm}
\caption{CPU and memory power consumption from profiling synthetic benchmarks on A57 using two cores. The labels in the x-axis are of the format <$f_{C}, f_{M}$>. }
\vspace{-1mm}
\label{fig:syn_power}
\end{figure}

\begin{figure*}[!t]
\centering
\vspace{-3mm}
\includegraphics[width=0.9\textwidth]{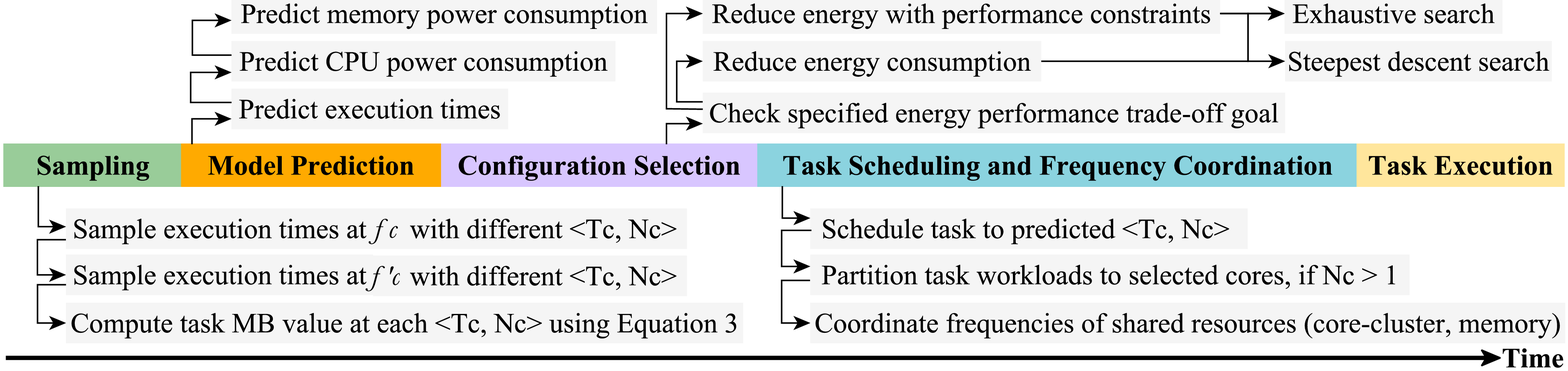}
\vspace{-9mm}
\caption{JOSS task scheduler timeline.}
\vspace{-5mm}
\label{fig:jossdesign}
\end{figure*}

\vspace{-1mm}
\subsection{Power Models}
The goal of the CPU and memory power models, used in JOSS, is to predict the impact of joint core and memory frequency scaling on CPU power and memory power consumed by a task, respectively.
We leverage the statistics collected from profiling synthetic benchmarks for building the CPU and memory power models, akin to the performance model, as shown in 
Figure~\ref{fig:modelbuilding}. 

\subsubsection{CPU Power Model}
The results from running synthetic benchmarks indicate that CPU power consumption is mainly dependent on core frequency and task characteristics (MB). 
\textcolor{black}{
For instance, Figure~\ref{fig_compow} indicates that CPU power consumption, when running three synthetic benchmarks (represented by different levels of MB) with various <$f_{C}, f_{M}$> settings (represented in x-axis) on Jetson TX2, shows negligible effects from memory frequency scaling.} 
Consequently, we build the CPU power model 
as shown in Equation~\ref{cpu_power}.
\textcolor{black}{We do not use voltage explicitly since it is strongly correlated with frequency in our evaluation platform and this enables us to reduce collinearity on the regression model.}
\vspace{-2mm}
\begin{equation}\label{cpu_power} \footnotesize
Power_{C} = \sum\limits_{0}^{1} \beta_{i}x_{i} + \sum\limits_{0}^{1} \beta_{ii}x_{i}^{2} + \beta_{01}x_{0}x_{1} + \varepsilon, \quad x_{i}=\{MB, f_{C}\}
\vspace{-2mm}
\end{equation}

\subsubsection{Memory Power Model}
Memory power is dependent on all three influential factors, i.e.~core frequency scaling, memory frequency scaling and task characteristics (MB).
Figure~\ref{fig_mempow} shows the impact of  $f_{C}, f_{M}$ and MB on memory power on Jetson TX2.
Consequently, we build the memory power model 
as shown in Equation~\ref{memory_power}.
\vspace{-2mm}
\begin{equation}\label{memory_power} \footnotesize
    Power_{M} = \sum\limits_{0}^{2} \beta_{i}x_{i} + \sum\limits_{0}^{2} \beta_{ii}x_{i}^{2} + \sum\limits_{i=0}^{1}\sum\limits_{k=i+1}^{2} \beta_{ik}x_{i}x_{k} + \varepsilon, \quad x_{i}=\{MB, f_{C}, f_{M}\}
    \vspace{-1mm}
\end{equation}

\subsubsection{Idle Power}
The total predicted power consumption for a task is the sum of dynamic power and idle power. 
Dynamic power consumed by a task is estimated using the models discussed previously. 
We measure idle CPU power and idle memory power during benchmarking when cores are switched on but are not actively executing computations and use the measured values as predictions. 
\textcolor{black}{We incorporate idle power characterization at different frequencies (voltages) in our models but do not consider temperature due to the small observed variations (<10 degrees) in operating temperature.}
Unlike dynamic power which is specific to each task, idle power is shared across all concurrently running tasks. 
We obtain information about the number of concurrently running tasks from the runtime (details in Section~\ref{scheduling}) and use that to attribute idle CPU and memory power proportionally among concurrently running tasks.

\textcolor{black}{\textbf{Modeling for different core type and number of cores:}}
The aforementioned models predict the impact of tuning two knobs <$f_{C}, f_{M}$> 
on performance, CPU and memory dynamic power consumption. 
However, when tasks execute on different core types and with different number of cores <$\mathrm{T_{C}}, \mathrm{N_{C}}$>,  MB values change due to the underlying core performance and workloads characteristics.
Consequently, the coefficients in the models for different <$\mathrm{T_{C}}, \mathrm{N_{C}}$> are distinct and we determine them 
via running the synthetic benchmarks at corresponding <$\mathrm{T_{C}}, \mathrm{N_{C}}$>.

\textcolor{black}{Our evaluation in Section~\ref{results} shows that the proposed models are accurate for determining the configuration for specified trade-off goals with low overhead during runtime.
We also evaluated the effectiveness of enhancing the performance and power models with higher degree coefficients but observed that it resulted in model overfitting and increased computation overheads without further improvement in prediction accuracy.
Note that the  profiling and the model building steps just need to be done once for a specific platform (e.g.~at install-time or boot-time), and do not impact the execution time of applications.}

\vspace{-1mm}
\section{JOSS Task Scheduler} \label{design}
JOSS task scheduler utilizes model predictions to take scheduling decisions and explore energy performance trade-offs. 
Figure \ref{fig:jossdesign} provides an overview of the scheduler's timeline. 
JOSS first samples task execution times to obtain MB values 
required for performance and power predictions.
Details regarding the sampling process and model invocation are presented in Section~\ref{SamplePrediction}.
Next, in Section~\ref{configselect}, we discuss how JOSS employs the predictions and identifies the best configuration for the four knobs to satisfy the energy performance trade-off goal.
In Section~\ref{scheduling}, we discuss the task scheduling process and frequency coordination approach applied to shared resources (i.e.~core-clusters and memory) \textcolor{black}{where the frequency throttling by concurrently running tasks could potentially lead to interference}. 

\vspace{-2mm}
\subsection{Runtime Sampling and Model Prediction} \label{SamplePrediction}
As discussed earlier, in Equation~\ref{MB}, power and performance models rely on MB values which are computed by sampling task execution times at two different core frequency settings $f_{C}$ and $f_{C}'$.
Furthermore, it is important to sample task execution with different <$\mathrm{T_{C}}, \mathrm{N_{C}}$> configurations since 
MB values vary with different core types and number of cores used to execute a task.
Consequently, JOSS samples task execution times, for different kernels, when running with different <$\mathrm{T_{C}}, \mathrm{N_{C}}$> configurations at both $f_{C}$ and $f_{C}'$ and then uses the MB values for predicting at different <$f_{C}, f_{M}$>.

JOSS performs online sampling (at the beginning of execution) for each kernel.
It leverages the observation that a typical kernel is invoked several times during application execution and that it is sufficient to sample a small fraction to estimate MB without introducing prohibitive overheads from online sampling.
The task scheduler initializes a separate performance look-up table, a CPU power look-up table and a memory power look-up table for storing the measured values and the predictions for each kernel. 

In a nutshell, the runtime sampling and model prediction phase operates as follows:
Firstly, JOSS samples the execution times of all kernels at different <$\mathrm{T_{C}}, \mathrm{N_{C}}$> at $f_{C}$. 
Once all kernels are sampled at $f_{C}$, JOSS then switches the cluster frequency to $f_{C}'$ and repeats the process. 
Note that the frequency transitions on different clusters are asynchronous, i.e.~sampling on one cluster can immediately transition from $f_{C}$ to $f_{C}'$ without waiting for the sampling completion on the other clusters. 
After sampling at $f_{C}'$, JOSS immediately computes the MB values at different <$\mathrm{T_{C}}, \mathrm{N_{C}}$> configurations. 
It then uses it along with performance and power models to populate the per-kernel look-up tables with predicted values. 
For the benchmarks we evaluate, our analysis shows that JOSS only spends 0.8\% of the total execution time, on average, in this phase.

\vspace{-2mm}
\subsection{Configuration Selection for Different Energy Performance Trade-off Goals} \label{configselect}
Once 
model predictions are complete, the scheduler transitions to configuration selection for each kernel. 
JOSS achieves the desired energy performance trade-off goal by optimizing the execution of each individual task.
For instance, JOSS reduces the total energy consumption by running each task with the lowest energy possible.
JOSS utilizes predictions to determine the configuration that satisfies the desired energy performance trade-off for each kernel. 
The approach for selecting the best configuration for each kernel is detailed later in this section. Successive invocations of the same kernel use the identified configuration without having to incur the overhead of configuration selection repeatedly. 
In this section, we investigate two different scenarios: reducing total energy consumption with 
and without performance constraints.

\vspace{-1mm}
\subsubsection{Reducing Total Energy Consumption} \label{energymini}
A simple approach for configuration selection is to exhaustively loop through 
all possible configurations and compare the estimated energy values to determine the configuration that consumes the least energy. 
However, as core counts and the number of available DVFS settings scale, such an approach can result in significant computation overheads during runtime.
To address this, we introduce a heuristic search algorithm based on the steepest descent method that can prune the large search space and identify the configuration with the least energy consumption with reduced overhead. 

\begin{figure}[!b]
\centering
\vspace{-6mm}
\includegraphics[width=0.85\columnwidth]{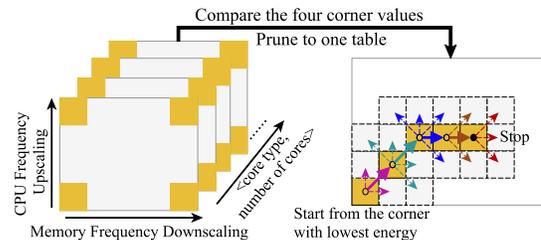}
\vspace{-5mm}
\caption{The steepest descent approach used in JOSS for pruning search space and configuration selection. 
}
\vspace{-2mm}
\label{fig:hillclimbing}
\end{figure}

Figure~\ref{fig:hillclimbing} illustrates the pruning process.
\textcolor{black}{First,} the algorithm computes the energy consumption of four corner configurations (representing combinations of the highest and the lowest CPU and memory frequency)
for each <$\mathrm{T_{C}}, \mathrm{N_{C}}$>.
\textcolor{black}{Second,} the algorithm compares the four corner values across different <$\mathrm{T_{C}}, \mathrm{N_{C}}$> to identify the <$\mathrm{T_{C}}, \mathrm{N_{C}}$> with the most number of lowest corner values.
This step confines the search space to a specific <$\mathrm{T_{C}}, \mathrm{N_{C}}$> table.
In the third step, the algorithm searches for the most energy-efficient joint DVFS setting <$f_{C}$, $f_{M}$> from this table.
This is accomplished by starting from the corner that has the least energy consumption, comparing the energy consumption of that configuration against all its immediate neighbours and repeating this immediate neighbour search process iteratively until it converges at a configuration with the least energy consumption.
The algorithm terminates once it detects that the energy value of the selected configuration is the lowest among all its immediate neighbors. 
We compare the overheads and the effectiveness of the two approaches in Section~\ref{overhead}.

\vspace{-1mm}
\subsubsection{Reducing Total Energy Consumption under Performance Constraints}
JOSS supports adding a performance constraint with respect to the configuration that provides the least energy consumption. 
The constraint is specified in the form of a performance speedup, which is used by the JOSS task scheduler to increase the performance of each individual task in the DAG. 
More specifically, for each task, the scheduler starts from the configuration that minimizes energy, then indexes into the corresponding performance look-up table and searches for configurations that satisfy the speedup constraint.
Both exhaustive search or steepest descent search can be used to locate the configuration with the lowest energy consumption. 
If no configuration meets the  performance constraint, the fastest configuration is selected.

\begin{table*}[!t]
\footnotesize 
\vspace{-2mm}
\caption{Evaluated Benchmarks}
\vspace{-8mm}
\begin{center}
\begin{tabular}{@{}p{1.9cm}p{0.35cm}p{9.1cm}p{3.5cm}p{1.7cm}@{}}\\ \toprule
\textbf{Benchmark} & \textbf{abbr.} & \textbf{Description} & \textbf{Input Size} & \textbf{Num. of Tasks}\\ \midrule
\textbf{Heat Diffusion}~\cite{XiTAO} & HD & Heat diffusion on a 2D grid using the iterative Jacobi stencil, which includes two kernels: Copy and Jacobi. We evaluate three problem sizes of different resolutions. & 2048(small),8192(big), 16384(huge) & 320032, 32032, 16032 \\ \midrule
\textbf{Dot Product}~\cite{XiTAO} & DP & Computing the sum of the products of two equal-length vectors, vectors are partitioned into blocks and computation of each block is marked as a single task, 100 iterations. & VectorSize 6400000, BlockSize 32000 & 20200 \\ \midrule
\textbf{Fibonacci}~\cite{BarcelonaBenchmark} & FB & Fibonacci numbers computed using recursion method. & Term 55, GrainSize 34 & 57314\\ \midrule
\textbf{Darknet-VGG-16 CNN}~\cite{VGG16} & VG & A 16-layered deep neural network that is typical of mobile and edge devices and implemented as a fork-join DAG, iteratively executed for 10 iterations. & 768$\times$576 RGB image, blocksize 64 & 5090 \\ \midrule
\textbf{Biomarker Infection}~\cite{BiomarkerAPP} & BI & A medical usecase for differentiating periprosthetic hip infection and aseptic hip prosthesis loosening. It computes the possible Biomarkers combinations to predict symptoms. & Sample Size 2 & 6217 \\ \midrule 
\textbf{Alya}~\cite{Alya} & AL & Alya is a high performance computational mechanics code to solve complex partial differential equations, and the parallelization strategy is based on mesh partitioning. & 200K CSR non-zeros & 47840 \\ \midrule 
\textbf{Sparse LU Factorization}~\cite{BarcelonaBenchmark} & SLU & Sparse matrix decomposition into the product of a lower and upper triangular matrix. It includes four kernels: LU0, FWD, BDIV and BMOD. & 64 blocks, BlockSize 512 & 11472 \\ \midrule 
\textbf{Matrix Multiplication}~\cite{XiTAO} & MM & A synthetic benchmark where each task computes A$\times$B=C, A and B are partitioned in N$\times$N tiles, N = input size. $dop$ is configurable. & 256$\times$256, 512$\times$512 & 10000, 2000\\ \midrule
\textbf{Matrix Copy}~\cite{XiTAO} & MC & A synthetic benchmark where each task reads and writes a large matrix, creating streaming behavior to access the main memory continuously. $dop$ is configurable. & 4096$\times$4096, 8192$\times$8192 & 20000, 10000  \\ \midrule
\textbf{Stencil}~\cite{XiTAO} & ST & A synthetic benchmark where each task repeatedly updates points on a multi-dimensional grid using the values at a set of neighboring points. $dop$ is configurable. & 512$\times$512, 2048$\times$2048 & 50000, 50000 \\ \bottomrule
\end{tabular}
\label{tab:benchmarks}
\vspace{-4mm}
\end{center}
\end{table*}

\vspace{-1mm}
\subsection{Task Scheduling and Frequency Coordination} \label{scheduling}
Once the task scheduler determines the configuration for a ready task (i.e.~input dependencies of the task are satisfied and can be scheduled for execution), it places the task in a work queue of a randomly selected core of the determined core type.
Note that the scheduler allows the task to be stolen by other cores of the same type, to maintain load balancing while also ensuring that the task runs on the most suitable core type identified in the previous phase. 
Task moldable execution ($\mathrm{N_{C}}$>1) is performed on multiple cores by dynamically partitioning the task workloads among cores of the same type.
Once a core finishes executing a task partition, it can continue fetching other tasks from its own work queue without waiting for partitions to finish 
on other cores.
The core that finishes executing the partition last, declares the completion of the task and wakes up the dependent tasks.

JOSS tracks the status of each core (i.e.~working or sleeping) to estimate instantaneous task concurrency. This is required to attribute the shared idle power among concurrently executing tasks. 
Furthermore, frequency throttling of the shared resources such as core-cluster and memory subsystem impacts concurrently executing tasks. 
Diverse frequency requirements for concurrent tasks can result in DVFS interference on shared resources and trigger DVFS serialization thereby introducing performance bottlenecks.
JOSS therefore adopts a simple averaging heuristic to balance the demands among the concurrent tasks when it detects that there is concurrency. 
JOSS averages the pre-determined frequency setting for the task with the current frequency setting of the shared resources.
We evaluated other heuristics such as \textit{min}, \textit{max}, weighted average,~etc.~and found arithmetic mean to perform the best.  

\textcolor{black}{\textbf{Fine-grained tasks:}} DVFS throttling overhead for fine-grained tasks where the execution times can be as small as a few microseconds is non-negligible.
Therefore, JOSS adopts the task coarsening algorithm proposed in the state-of-the-art~\cite{STEER/SBACPAD2022}, which first determines the <$\mathrm{T_{C}}, \mathrm{N_{C}}$> without any frequency throttling and then attempts to search for more tasks of the same type from the work queues of the selected core type in a round-robin manner. 
Once a sufficient number of fine-grained tasks of the same type is found, JOSS searches for the best joint <$f_{C}, f_{M}$> setting that satisfies the trade-off under the determined <$\mathrm{T_{C}}, \mathrm{N_{C}}$>.

\vspace{-1mm}
\section{Experimental Methodology} \label{setup}

\subsection{Experimental Platform} \label{platform}
We use the NVIDIA Jetson TX2 development board in our evaluation~\cite{JetsonTX2Module}. 
It is an asymmetric platform that features two CPU clusters: Denver and A57. 
The Denver cluster comprises a high-performance dual-core NVIDIA Denver CPU, while the A57 cluster comprises a comparatively lower performance quad-core ARM CPU. 
Both clusters support the same range of operating core frequencies.
The two clusters can be operated at different frequencies but all the cores in the same cluster must operate at the same frequency.
The choice of using the NVIDIA Jetson TX2 is also motivated by its support for EMC frequency scaling for the memory controller (EMC/MC) and DRAM (LPDDR4). 
Existing systems with support for memory DVFS typically support frequency scaling in the memory controller, the DDRIO and the DRAM device while only supporting voltage scaling in the memory controller due to design challenges associated with operating the DRAM array at multiple voltages \cite{jawad}.
The integrated \texttt{INA3221} power sensor is used to sample the power consumption of CPU and memory subsystem.
Power samples obtained every 5 milliseconds are used to compute CPU and memory energy consumption, which is then accumulated throughout the duration of application execution.
\textcolor{black}{The Linux governor is set as \texttt{userspace} to enable CPU frequency scaling.}
The CPU and memory frequency are set at the highest, i.e.~2.04GHz for both clusters and 1.87GHz for memory, before executing a benchmark.
\textcolor{black}{The Linux kernel version is 4.9.253-tegra and the compiler version is g++ 7.5.0.}
We repeat each experiment 10 times and report the arithmetic average.

\vspace{-2mm}
\subsection{Evaluated Benchmarks and Schedulers} \label{schedulers}
We evaluate JOSS using ten 
\textcolor{black}{benchmarks} from the Edge and HPC domains.
Table~\ref{tab:benchmarks} provides additional details.
These benchmarks comprise a different number of kernels \textcolor{black}{(i.e.~task types)} and 
exploit parallelism by invoking multiple instances of them. 

We evaluate the effectiveness of JOSS by comparing it to multiple state-of-the-art task-based schedulers.
Both JOSS and the evaluated schedulers below are implemented on top of XiTAO~\cite{XiTAO}.


(1) \textit{GRWS} (Greedy Random Work Stealing) is a widely used baseline scheduler for task-based applications~\cite{frigo-cilk5,tbb_scheduler,openmp50-api}, which
attempts to keep idle cores busy through task stealing. 
GRWS does not leverage DVFS knobs and each task only runs on a single core. 

(2) \textit{ERASE}~\cite{Jing_ERASE} employs an online history-based performance model and an offline categorized CPU power model to determine the configuration <$\mathrm{T_{C}}, \mathrm{N_{C}}$> that reduces CPU energy consumption without relying on explicit DVFS changes. 


(3) \textit{Aequitas}~\cite{HarisRibic-AEQUITAS} is a heuristic-based scheduler that extends HERMES~\cite{HarisRibic-WorkStealing/PercoreDVFS/Tempo}. 
It first determines the core frequency for running each task based on task thief-victim relations (slow down the thief cores) and the size of the work queues.
On core-clustered platforms, it lets each active core within a cluster tune the cluster frequency for a short interval (1s) in a round-robin time-slicing manner. 
It does not leverage the memory DVFS knob and moldable execution.

(4) \textit{STEER}~\cite{STEER/SBACPAD2022} is a model-based scheduler, which exploits the task characteristics and available CPU DVFS knob. 
It utilizes a performance model and a CPU power model to identify the configuration <$\mathrm{T_{C}}, \mathrm{N_{C}}, f_{C}$> for each task that consumes the least CPU energy.
STEER does not leverage the memory DVFS knob.
\begin{figure*}[!t]
\centering
\vspace{-4mm}
\includegraphics[width=\textwidth]{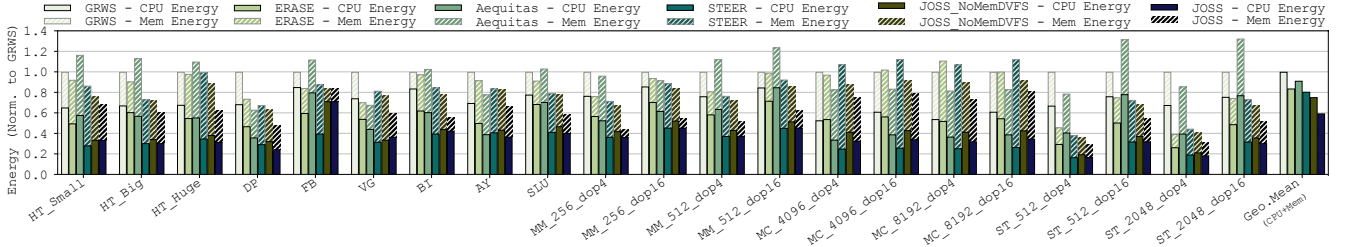}
\vspace{-9mm}
\caption{Total energy consumption of evaluated benchmarks when using GRWS, Aequitas, ERASE, STEER and JOSS. 
All energy values are normalized to the total energy of the baseline GRWS, therefore, lower is better.}
\vspace{-5mm}
\label{fig:state-of-the-art}
\end{figure*}

\vspace{-1mm}
\section{Evaluation} \label{results}
We evaluate JOSS under two scenarios targeting different energy performance trade-offs: 
(i) in Section~\ref{energy} we evaluate the effectiveness of JOSS at reducing the total energy consumption by comparing it to several state-of-the-art schedulers; 
(ii) in Section~\ref{energy_perf} we evaluate the ability of JOSS for reducing the total energy consumption with user specified performance constraints with respect to (i).
Finally, we analyze the prediction accuracy of three proposed models in Section~\ref{accuracy} and present overhead analysis in Section~\ref{overhead}.

\vspace{-2mm}
\subsection{Reducing Total Energy Consumption} \label{energy}
Figure~\ref{fig:state-of-the-art} compares the total energy consumption (incl.~CPU energy and memory energy) when using GRWS, ERASE, Aequitas, STEER and JOSS across different benchmarks. 
We also include a new datapoint, JOSS\_NoMemDVFS, where JOSS is employed for reducing the total energy consumption without leveraging the memory DVFS knob ($f_{M}$ is fixed at max. value). 
This is included to understand the impact of JOSS on asymmetric platforms, which support CPU DVFS but lack support for memory DVFS.
For the supported benchmarks, we evaluate different \textcolor{black}{task granularity} and task DAG parallelism ($dop$) settings.
This enables us to evaluate the effectiveness of the schedulers across a broad spectrum of task DAGs. 

Overall, the results show that JOSS consumes the least energy across all benchmarks compared to the evaluated schedulers. 
Specifically, JOSS achieves 40.7\% energy reduction, on average, compared to the baseline GRWS, while STEER, Aequitas and ERASE achieve 19.5\%, 8.7\% and 16.3\% average reduction compared to the baseline respectively. 
These results demonstrate that JOSS achieves an additional 21.2\% energy reduction compared to STEER (the best among the state-of-the-art).
Even in the absence of memory DVFS knob, JOSS\_NoMemDVFS achieves a 24.8\% reduction in energy consumption compared to GRWS, which is still an improvement over the state-of-the-art (e.g.~5.2\% additional savings than STEER). 
This emphasizes the importance of taking the total energy consumption into account even when the memory DVFS knob is unavailable.  

We analyze the effectiveness of JOSS using SparseLU (specifically the BMOD kernel) as an example. BMOD kernel accounts for 91\% of the total number of tasks in SparseLU. 
With GRWS, 63\% of BMOD tasks execute on the high-performance Denver cores while 37\% of tasks execute on the relatively low-performance A57 cores.  
Although executing on a single Denver core is 3.4$\times$ faster than an A57 core, a reasonable fraction end up executing on the A57 cores, since the four A57 cores end up stealing more tasks from Denver queues. 
With ERASE, the CPU energy estimates obtained using performance and CPU power models indicate that running BMOD tasks on two Denver cores consumes less CPU energy since it can achieve linear speedup without doubling the CPU power consumption. 
Thus, ERASE reduces the CPU energy compared to GRWS. 
Aequitas relies on task stealing relations and the work queue size to select the core frequency and each active core tunes the cluster frequency for a short interval. 
A57 cores steal more BMOD tasks from the Denver cores, which make A57 become thief cores and get more workloads.
Therefore, it ends up both slowing down and speeding up A57 cluster frequency for brief periods during execution (38\% tasks executing on Denver and 62\% executing on A57). 
With CPU frequency throttling, Aequitas reduces CPU energy but increases memory energy consumption in comparison to GRWS and ERASE due to the performance slowdown. 
STEER further reduces the CPU energy consumption by identifying the configuration <Denver, 2, 1.11GHz> for BMOD tasks. 
STEER however does not take  memory energy consumption into account. Consequently, the performance slowdown from throttling CPU frequency setting results in higher memory energy consumption. 

In contrast to STEER, JOSS\_NoMemDVFS aims to reduce the total energy consumption. 
It utilizes three proposed models in JOSS to predict the CPU energy together with memory energy when only throttling the core frequency while memory frequency is fixed as the maximum.  
JOSS\_NoMemDVFS selects <Denver, 2, 1.57GHz> as the configuration for reducing energy consumption. 
Running at 1.57GHz increases the CPU energy consumption compared to STEER. 
However, it also ends up reducing more memory energy consumption because of the performance improvement achieved from higher CPU frequency. 
JOSS leverages the memory frequency knob to further reduce the total energy consumption by identifying the configuration of <Denver, 2, 1.11GHz, 0.8GHz> for BMOD tasks.
Since BMOD kernel is compute-intensive when running on two Denver cores (MB is estimated to be 1\%), running with lower memory frequency does not have much impact on execution time of the tasks and leads to lower memory energy consumption. 

\begin{figure*}[!t]
\centering
\vspace{-5mm}
\includegraphics[width=\textwidth]{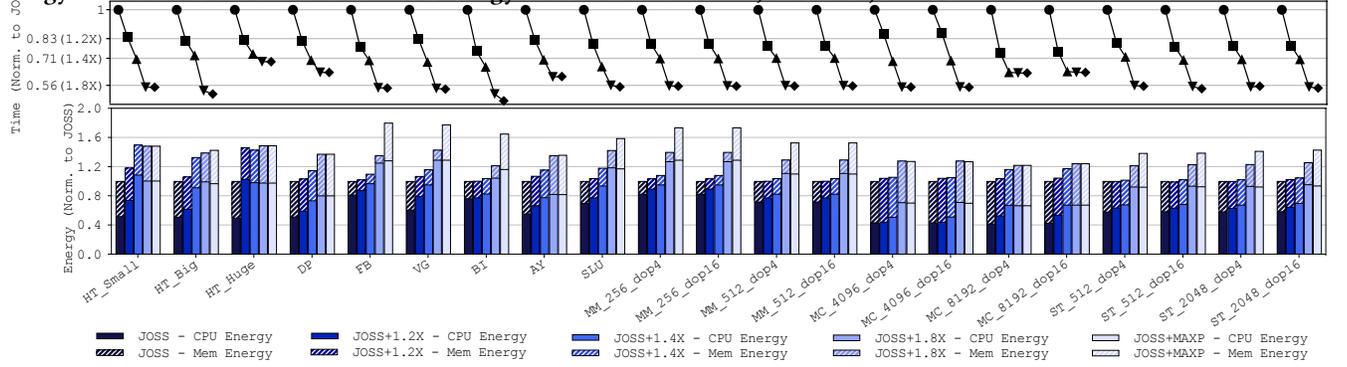}
\vspace{-8mm}
\caption{Energy consumption (bottom) and execution time (top) when targeting energy reduction under performance constraints. Performance and energy values are normalized to JOSS without performance constraints.}
\vspace{-5mm}
\label{fig:epto_joss}
\end{figure*}

\begin{figure}[!t]
\centering
\includegraphics[width=0.7\columnwidth]{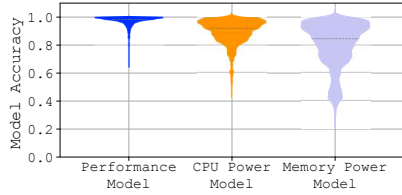}
\vspace{-5mm}
\caption{Model prediction accuracy of three proposed models in JOSS. Dotted lines represent the medians.}
\vspace{-6mm}
\label{fig:modelaccuracy}
\end{figure}

\vspace{-2mm}
\subsection{Reducing Total Energy under Performance Constraints} \label{energy_perf}
Figure~\ref{fig:epto_joss} shows results of 
JOSS reducing the total energy consumption while attempting to satisfy user specified performance constraints.  
In this experiment, we test three performance targets (speedups of 1.2$\times$, 1.4$\times$ and 1.8$\times$ with respect to JOSS targeting energy reduction solely), in addition to MAXP where JOSS maximizes individual task performance without considering energy. 

The top part of the figure shows the execution time of each configuration, along with the performance target. 
Overall, the results across benchmarks show that JOSS can achieve 1.2$\times$, 1.4$\times$ and 1.8$\times$ speedups at the additional cost of 6\%, 13\% and 32\% increase in energy consumption over JOSS without performance constraints. 
In a few cases, the ability to achieve the desired trade-off targets is impacted by the accuracy of the prediction models.
For example, in the case of MC\_4096, JOSS slightly misses (by 3\%) the deadline of 1.2$\times$ speedup due to the inaccuracy in model predictions.
\textcolor{black}{The average prediction error of performance, CPU power and memory power models in this case are 9.2\%, 13.8\% and 18.9\%, respectively.}
Furthermore, in benchmarks with high degree of memory intensity, JOSS does not achieve 1.8$\times$ speedup even when executing with maximum <$f_{C}, f_{M}$>, despite significant increase in total energy consumption. Ultimately, task performance is limited by processor capabilities, such as peak FLOPS and memory bandwidth, restricting JOSS' ability to reach a performance target. 

\vspace{-2mm}
\subsection{Model Accuracy} \label{accuracy}
We analyze the accuracy of the performance, CPU power and memory power models proposed in JOSS.
We compute accuracy using the formula: $accuracy = 1 - \frac{Absolute(real - prediction)}{real}$.
We report the arithmetic average numbers across all evaluated benchmarks. 
The real values are collected through running each benchmark on all possible configurations for the four knobs.
Figure~\ref{fig:modelaccuracy} presents the prediction accuracy distribution of the three models across the evaluated benchmarks. 
The results show that the performance model achieves 97\% accuracy on average (the median is 98.3\% shown as the dotted line in Figure~\ref{fig:modelaccuracy}), the CPU power model achieves 90\% accuracy on  average (the median is 91.8\%), while the memory power model achieves 80\% accuracy on average (the median is 84.6\%).

\vspace{-2mm}
\subsection{Overhead Analysis}\label{overhead}
To enable performance and power consumption prediction, JOSS implements three look-up tables per kernel for storing the measured and predicted execution times, CPU power and memory power consumption. 
Consider a platform with $N$ cores in total, $M$ asymmetric core-clusters such that each cluster comprises $\frac{N}{M}$ cores, the possible number of cores that can be used for each task equals $\log\frac{N}{M}$.
Assume the numbers of available core frequency and memory frequency settings are $Nf_{C}$ and $Nf_{M}$. 
The storage overhead for three look-up tables for each kernel in JOSS is $3 \times M \times \log\frac{N}{M} \times Nf_{C} \times Nf_{M}$. 

Next we compare the overheads of using steepest descent search and exhaustive search
(details in Section~\ref{energymini}). 
The results on Jetson TX2 show that using steepest descent search reduces timing overheads by 70\% on average, compared to exhaustive search across all evaluated benchmarks, due to the significant reduction in the number of comparisons.
Our evaluation also shows that 
configurations using selected steepest descent search achieves 97\% energy reduction relative to the configurations selected using exhaustive search. 
On larger platforms, using the steepest descent search is expected to reduce timing overheads even further. 
\vspace{-1mm}
\section{Related Work}
Existing works can broadly classified into three categories: those focusing on CPU energy reduction, memory energy reduction and total (CPU and memory) energy reduction. 

\textit{CPU energy on architectures with per-core DVFS:} Acun et al.~\cite{PerCore/Kernel/Granularity/IGSC2019} employ per-core DVFS and adopt an online history-based approach for performance and CPU power predictions by executing with every possible frequency.
HERMES~\cite{HarisRibic-WorkStealing/PercoreDVFS/Tempo} proposes workpath-sensitive and workload-sensitive algorithms for per-core DVFS in a work stealing runtime.
It slows down thief cores and selects appropriate frequencies based on workload sizes. 
CATA~\cite{EmilioCastillo-IPDPS2016-CATA/Criticality/EDP} 
dynamically tunes the frequency based on incoming task criticality and the available power budget at the moment.
AAWS~\cite{Christopher-ISCA2016-Asymmetric/PercoreDVFS/Parallelism} targets for 
work stealing runtime on asymmetric platforms and proposes work-pacing, work-sprinting and work-mugging strategies (that require hardware support)
by detecting the parallel slackness.

\textit{CPU energy on architectures with cluster-based DVFS: }
\textcolor{black}{Besides Aequitas~\cite{HarisRibic-AEQUITAS}, ERASE~\cite{Jing_ERASE} and STEER~\cite{STEER/SBACPAD2022}, discussed in Section~\ref{schedulers},}
CHRT~\cite{HMP/DVFS/Criticality/Ratio/Prediction/TPDS2020} is a phase-based scheduler that predicts task placement, cluster frequency, and number of cores for each execution phase. It uses an offline model where the online phases map to the model and takes the recorded configuration as the prediction.

\textit{Memory energy: }
MemScale~\cite{MemScale/ASPLOS2011} targets energy consumption in the memory subsystem. They leverage dynamic profiling, performance and power modeling to guide DVFS of memory controller and frequency scaling of memory channels and DRAM. 
David et al.~\cite{MemDVFS/ICAC2011} propose an intuitive algorithm that detects memory bandwidth utilization for tuning DVFS of memory subsystem. Both proposals are however tailored for multi-programmed workloads.

\textit{CPU+Memory energy: }
Sundriyal et al~\cite{JointDVFS/Supercomputing16} targets minimizing the power consumption of a system given the performance loss tolerance.
They propose performance and power models using PMCs to determine the best joint frequency setting in a time window-based manner for the entire application. 
CoScale~\cite{Coscale/MICRO2012} is epoch-based framework for multi-programmed workloads. They first collect PMCs for model prediction and then search for the best frequency pair using gradient-descent. 

Another line of related work~\cite{MaxPerf/Powerlimit/ICS13} targets maximizing performance of data-parallel applications under power constraints instead of reducing energy consumption.

\section{Conclusion}
We propose JOSS, a runtime scheduling framework that can both reduce energy consumption and explore various energy performance trade-offs for task-based parallel applications. 
Overall, JOSS achieves the goals set out via optimizing the execution of each task in the application. 
JOSS comprises a performance model, a CPU power model, a memory power model and a task scheduler.
It utilizes the three models to
predict the execution time and power consumption for each task when running with different configurations for the four knobs (i.e.~core type, number of cores, core frequency scaling and memory frequency scaling). 
In contrast to existing works, JOSS manages to achieve higher energy savings by considering the impact of memory energy consumption, in addition to core asymmetry, CPU DVFS and task characteristics, and through the use of memory DVFS as a tunable knob. 
Our evaluation shows that JOSS achieves an additional 21.2\% energy reduction on average compared to the state-of-the-art. 
Even in the absence of memory DVFS knob, JOSS can still save 5.2\% additional energy. 
Furthermore, it is capable of reducing the total energy while still satisfying the performance constraints specified. 
\textcolor{black}{We hope that these results together with related papers that demonstrate benefits of memory scaling will encourage widespread adoption of memory DVFS knob as an additional avenue for improving energy efficiency.}

\bibliographystyle{ACM-Reference-Format} 
\bibliography{Ref,jrlib-miquel}


\begin{thebibliography}{46}


\ifx \showCODEN    \undefined \def \showCODEN     #1{\unskip}     \fi
\ifx \showDOI      \undefined \def \showDOI       #1{#1}\fi
\ifx \showISBNx    \undefined \def \showISBNx     #1{\unskip}     \fi
\ifx \showISBNxiii \undefined \def \showISBNxiii  #1{\unskip}     \fi
\ifx \showISSN     \undefined \def \showISSN      #1{\unskip}     \fi
\ifx \showLCCN     \undefined \def \showLCCN      #1{\unskip}     \fi
\ifx \shownote     \undefined \def \shownote      #1{#1}          \fi
\ifx \showarticletitle \undefined \def \showarticletitle #1{#1}   \fi
\ifx \showURL      \undefined \def \showURL       {\relax}        \fi
\providecommand\bibfield[2]{#2}
\providecommand\bibinfo[2]{#2}
\providecommand\natexlab[1]{#1}
\providecommand\showeprint[2][]{arXiv:#2}

\bibitem[\protect\citeauthoryear{??}{Sta}{2014}]%
        {StarPU}
 \bibinfo{year}{2014}\natexlab{}.
\newblock \bibinfo{title}{Documentation of StarPU}.
\newblock
  \bibinfo{howpublished}{\url{https://files.inria.fr/starpu/doc/starpu.pdf}}.
\newblock


\bibitem[\protect\citeauthoryear{??}{odr}{2015}]%
        {odroid-xu4}
 \bibinfo{year}{2015}\natexlab{}.
\newblock \bibinfo{title}{{ODROID XU4}}.
\newblock
  \bibinfo{howpublished}{\url{https://magazine.odroid.com/wp-content/uploads/odroid-xu4-user-manual.pdf}}.
\newblock


\bibitem[\protect\citeauthoryear{??}{Jet}{2017}]%
        {JetsonTX2Module}
 \bibinfo{year}{2017}\natexlab{}.
\newblock \bibinfo{title}{Jetson TX2 Module}.
\newblock
  \bibinfo{howpublished}{\url{https://developer.nvidia.com/embedded/jetson-tx2}}.
\newblock


\bibitem[\protect\citeauthoryear{??}{XiT}{2018}]%
        {XiTAO}
 \bibinfo{year}{2018}\natexlab{}.
\newblock \bibinfo{title}{{XiTAO} Runtime}.
\newblock
  \bibinfo{howpublished}{\url{https://github.com/CHART-Team/xitao.git}}.
\newblock


\bibitem[\protect\citeauthoryear{??}{Mem}{2019}]%
        {MemSpeed}
 \bibinfo{year}{2019}\natexlab{}.
\newblock \bibinfo{title}{DDR5/4/3/2: How Memory Density and Speed Increased
  with each Generation of DDR}.
\newblock
  \bibinfo{howpublished}{\url{https://blogs.synopsys.com/vip-central/2019/02/27/ddr5-4-3-2-how-memory-density-and-speed-increased-with-each-generation-of-ddr/}}.
\newblock


\bibitem[\protect\citeauthoryear{??}{arm}{2020}]%
        {arm-big-little}
 \bibinfo{year}{2020}\natexlab{}.
\newblock \bibinfo{title}{ARM BIG.LITTLE}.
\newblock
  \bibinfo{howpublished}{\url{https://www.arm.com/why-arm/technologies/big-little}}.
\newblock


\bibitem[\protect\citeauthoryear{??}{Bio}{2020}]%
        {BiomarkerAPP}
 \bibinfo{year}{2020}\natexlab{}.
\newblock \bibinfo{title}{Biomarker Discovery}.
\newblock
  \bibinfo{howpublished}{\url{https://legato-project.eu/use-cases/healthcare}}.
\newblock


\bibitem[\protect\citeauthoryear{??}{App}{2022}]%
        {AppleA16}
 \bibinfo{year}{2022}\natexlab{}.
\newblock \bibinfo{title}{Apple A16 Bionic}.
\newblock
  \bibinfo{howpublished}{\url{https://en.wikipedia.org/wiki/Apple_A16}}.
\newblock


\bibitem[\protect\citeauthoryear{Acun, Chandrasekar, and Kale}{Acun
  et~al\mbox{.}}{2019}]%
        {PerCore/Kernel/Granularity/IGSC2019}
\bibfield{author}{\bibinfo{person}{Bilge Acun}, \bibinfo{person}{Kavitha
  Chandrasekar}, {and} \bibinfo{person}{Laxmikant~V. Kale}.}
  \bibinfo{year}{2019}\natexlab{}.
\newblock \showarticletitle{Fine-Grained Energy Efficiency Using Per-Core DVFS
  with an Adaptive Runtime System}. In \bibinfo{booktitle}{\emph{2019 Tenth
  International Green and Sustainable Computing Conference (IGSC)}}.
\newblock


\bibitem[\protect\citeauthoryear{{Castillo}, {Moreto}, {Casas}, {Alvarez},
  {Vallejo}, {Chronaki}, {Badia}, {Bosque}, {Beivide}, {Ayguade}, {Labarta},
  and {Valero}}{{Castillo} et~al\mbox{.}}{2016}]%
        {EmilioCastillo-IPDPS2016-CATA/Criticality/EDP}
\bibfield{author}{\bibinfo{person}{E. {Castillo}}, \bibinfo{person}{M.
  {Moreto}}, \bibinfo{person}{M. {Casas}}, \bibinfo{person}{L. {Alvarez}},
  \bibinfo{person}{E. {Vallejo}}, \bibinfo{person}{K. {Chronaki}},
  \bibinfo{person}{R. {Badia}}, \bibinfo{person}{J.~L. {Bosque}},
  \bibinfo{person}{R. {Beivide}}, \bibinfo{person}{E. {Ayguade}},
  \bibinfo{person}{J. {Labarta}}, {and} \bibinfo{person}{M. {Valero}}.}
  \bibinfo{year}{2016}\natexlab{}.
\newblock \showarticletitle{CATA: Criticality Aware Task Acceleration for
  Multicore Processors}. In \bibinfo{booktitle}{\emph{2016 IEEE International
  Parallel and Distributed Processing Symposium (IPDPS)}}.
\newblock


\bibitem[\protect\citeauthoryear{Chen, Manivannan, Abduljabbar, and
  Peric\`{a}s}{Chen et~al\mbox{.}}{2022a}]%
        {Jing_ERASE}
\bibfield{author}{\bibinfo{person}{Jing Chen}, \bibinfo{person}{Madhavan
  Manivannan}, \bibinfo{person}{Mustafa Abduljabbar}, {and}
  \bibinfo{person}{Miquel Peric\`{a}s}.} \bibinfo{year}{2022}\natexlab{a}.
\newblock \showarticletitle{{ERASE}: Energy Efficient Task Mapping and Resource
  Management for Work Stealing Runtimes}.
\newblock \bibinfo{journal}{\emph{ACM Trans. Archit. Code Optim.}}
  (\bibinfo{date}{mar} \bibinfo{year}{2022}).
\newblock


\bibitem[\protect\citeauthoryear{Chen, Manivannan, Goel, Abduljabbar, and
  Pericàs}{Chen et~al\mbox{.}}{2022b}]%
        {STEER/SBACPAD2022}
\bibfield{author}{\bibinfo{person}{Jing Chen}, \bibinfo{person}{Madhavan
  Manivannan}, \bibinfo{person}{Bhavishya Goel}, \bibinfo{person}{Mustafa
  Abduljabbar}, {and} \bibinfo{person}{Miquel Pericàs}.}
  \bibinfo{year}{2022}\natexlab{b}.
\newblock \showarticletitle{STEER: Asymmetry-aware Energy Efficient Task
  Scheduler for Cluster-based Multicore Architectures}. In
  \bibinfo{booktitle}{\emph{2022 IEEE 34th International Symposium on Computer
  Architecture and High Performance Computing (SBAC-PAD)}}.
\newblock


\bibitem[\protect\citeauthoryear{Contreras and Martonosi}{Contreras and
  Martonosi}{2008}]%
        {tbb_scheduler}
\bibfield{author}{\bibinfo{person}{Gilberto Contreras} {and}
  \bibinfo{person}{Margaret Martonosi}.} \bibinfo{year}{2008}\natexlab{}.
\newblock \showarticletitle{Characterizing and improving the performance of
  intel threading building blocks}. In \bibinfo{booktitle}{\emph{2008 IEEE
  International Symposium on Workload Characterization}}. IEEE,
  \bibinfo{pages}{57--66}.
\newblock


\bibitem[\protect\citeauthoryear{Coutinho~Demetrios, De~Sensi, Lorenzon,
  Georgiou, Nunez-Yanez, Eder, and Xavier-de Souza}{Coutinho~Demetrios
  et~al\mbox{.}}{2020}]%
        {Tradeoff/hetero/model/Energy2020}
\bibfield{author}{\bibinfo{person}{AM Coutinho~Demetrios},
  \bibinfo{person}{Daniele De~Sensi}, \bibinfo{person}{Arthur~Francisco
  Lorenzon}, \bibinfo{person}{Kyriakos Georgiou}, \bibinfo{person}{Jose
  Nunez-Yanez}, \bibinfo{person}{Kerstin Eder}, {and} \bibinfo{person}{Samuel
  Xavier-de Souza}.} \bibinfo{year}{2020}\natexlab{}.
\newblock \showarticletitle{Performance and energy trade-offs for parallel
  applications on heterogeneous multi-processing systems}.
\newblock \bibinfo{journal}{\emph{Energies}} \bibinfo{volume}{13},
  \bibinfo{number}{9} (\bibinfo{year}{2020}), \bibinfo{pages}{2409}.
\newblock


\bibitem[\protect\citeauthoryear{Das, Werner, Antonakakis, Polychronakis, and
  Monrose}{Das et~al\mbox{.}}{2019}]%
        {PMC_pitfall_SP2019}
\bibfield{author}{\bibinfo{person}{Sanjeev Das}, \bibinfo{person}{Jan Werner},
  \bibinfo{person}{Manos Antonakakis}, \bibinfo{person}{Michalis
  Polychronakis}, {and} \bibinfo{person}{Fabian Monrose}.}
  \bibinfo{year}{2019}\natexlab{}.
\newblock \showarticletitle{SoK: The Challenges, Pitfalls, and Perils of Using
  Hardware Performance Counters for Security}. In
  \bibinfo{booktitle}{\emph{2019 IEEE S\&P}}.
\newblock


\bibitem[\protect\citeauthoryear{David, Fallin, Gorbatov, Hanebutte, and
  Mutlu}{David et~al\mbox{.}}{2011}]%
        {MemDVFS/ICAC2011}
\bibfield{author}{\bibinfo{person}{Howard David}, \bibinfo{person}{Chris
  Fallin}, \bibinfo{person}{Eugene Gorbatov}, \bibinfo{person}{Ulf~R.
  Hanebutte}, {and} \bibinfo{person}{Onur Mutlu}.}
  \bibinfo{year}{2011}\natexlab{}.
\newblock \showarticletitle{Memory Power Management via Dynamic
  Voltage/Frequency Scaling}. In \bibinfo{booktitle}{\emph{Proceedings of the
  8th ACM International Conference on Autonomic Computing}}
  \emph{(\bibinfo{series}{ICAC '11})}. \bibinfo{pages}{31–40}.
\newblock


\bibitem[\protect\citeauthoryear{Deng, Meisner, Bhattacharjee, Wenisch, and
  Bianchini}{Deng et~al\mbox{.}}{2012}]%
        {Coscale/MICRO2012}
\bibfield{author}{\bibinfo{person}{Qingyuan Deng}, \bibinfo{person}{David
  Meisner}, \bibinfo{person}{Abhishek Bhattacharjee},
  \bibinfo{person}{Thomas~F. Wenisch}, {and} \bibinfo{person}{Ricardo
  Bianchini}.} \bibinfo{year}{2012}\natexlab{}.
\newblock \showarticletitle{CoScale: Coordinating CPU and Memory System DVFS in
  Server Systems}. In \bibinfo{booktitle}{\emph{2012 45th Annual IEEE/ACM
  International Symposium on Microarchitecture}}.
\newblock


\bibitem[\protect\citeauthoryear{Deng, Meisner, Ramos, Wenisch, and
  Bianchini}{Deng et~al\mbox{.}}{2011}]%
        {MemScale/ASPLOS2011}
\bibfield{author}{\bibinfo{person}{Qingyuan Deng}, \bibinfo{person}{David
  Meisner}, \bibinfo{person}{Luiz Ramos}, \bibinfo{person}{Thomas~F. Wenisch},
  {and} \bibinfo{person}{Ricardo Bianchini}.} \bibinfo{year}{2011}\natexlab{}.
\newblock \showarticletitle{MemScale: Active Low-Power Modes for Main Memory}.
  In \bibinfo{booktitle}{\emph{Proceedings of the Sixteenth International
  Conference on Architectural Support for Programming Languages and Operating
  Systems}}.
\newblock


\bibitem[\protect\citeauthoryear{Duran, Teruel, Ferrer, Bofill, and
  Parra}{Duran et~al\mbox{.}}{2009a}]%
        {BarcelonaBenchmark}
\bibfield{author}{\bibinfo{person}{Alejandro Duran}, \bibinfo{person}{Xavier
  Teruel}, \bibinfo{person}{Roger Ferrer}, \bibinfo{person}{Xavier Bofill},
  {and} \bibinfo{person}{Eduard Parra}.} \bibinfo{year}{2009}\natexlab{a}.
\newblock \showarticletitle{Barcelona OpenMP Tasks Suite: A Set of Benchmarks
  Targeting the Exploitation of Task Parallelism in OpenMP}.
\newblock \bibinfo{journal}{\emph{Proceedings of the International Conference
  on Parallel Processing}} (\bibinfo{date}{09} \bibinfo{year}{2009}).
\newblock


\bibitem[\protect\citeauthoryear{Duran, Teruel, Ferrer, Martorell, and
  Ayguade}{Duran et~al\mbox{.}}{2009b}]%
        {BOTS/ICPP2009}
\bibfield{author}{\bibinfo{person}{Alejandro Duran}, \bibinfo{person}{Xavier
  Teruel}, \bibinfo{person}{Roger Ferrer}, \bibinfo{person}{Xavier Martorell},
  {and} \bibinfo{person}{Eduard Ayguade}.} \bibinfo{year}{2009}\natexlab{b}.
\newblock \showarticletitle{Barcelona openmp tasks suite: A set of benchmarks
  targeting the exploitation of task parallelism in openmp}. In
  \bibinfo{booktitle}{\emph{2009 international conference on parallel
  processing}}. IEEE, \bibinfo{pages}{124--131}.
\newblock


\bibitem[\protect\citeauthoryear{Endrei, Jin, Dinh, Abramson, Poxon, DeRose,
  and de~Supinski}{Endrei et~al\mbox{.}}{2018}]%
        {Modeling/Tradeoff/SC2018}
\bibfield{author}{\bibinfo{person}{Mark Endrei}, \bibinfo{person}{Chao Jin},
  \bibinfo{person}{Minh~Ngoc Dinh}, \bibinfo{person}{David Abramson},
  \bibinfo{person}{Heidi Poxon}, \bibinfo{person}{Luiz DeRose}, {and}
  \bibinfo{person}{Bronis~R. de Supinski}.} \bibinfo{year}{2018}\natexlab{}.
\newblock \showarticletitle{Energy Efficiency Modeling of Parallel
  Applications}. In \bibinfo{booktitle}{\emph{SC18: International Conference
  for High Performance Computing, Networking, Storage and Analysis}}.
\newblock


\bibitem[\protect\citeauthoryear{Frigo, Leiserson, and Randall}{Frigo
  et~al\mbox{.}}{1998}]%
        {frigo-cilk5}
\bibfield{author}{\bibinfo{person}{Matteo Frigo}, \bibinfo{person}{Charles~E.
  Leiserson}, {and} \bibinfo{person}{Keith~H. Randall}.}
  \bibinfo{year}{1998}\natexlab{}.
\newblock \showarticletitle{{The Implementation of the Cilk-5 Multithreaded
  Language}}, In \bibinfo{booktitle}{{Proceedings of SIGPLAN 1998}}.
\newblock \bibinfo{journal}{\emph{SIGPLAN}}.
\newblock


\bibitem[\protect\citeauthoryear{Goel}{Goel}{2016}]%
        {goel2016measurement}
\bibfield{author}{\bibinfo{person}{Bhavishya Goel}.}
  \bibinfo{year}{2016}\natexlab{}.
\newblock \bibinfo{booktitle}{\emph{Measurement, Modeling, and Characterization
  for Energy-efficient Computing}}.
\newblock \bibinfo{publisher}{Chalmers University of Technology}.
\newblock


\bibitem[\protect\citeauthoryear{Guillaume and Mariano}{Guillaume and
  Mariano}{[n.d.]}]%
        {Alya}
\bibfield{author}{\bibinfo{person}{Houzeaux Guillaume} {and}
  \bibinfo{person}{Vazquez Mariano}.} \bibinfo{year}{[n.d.]}\natexlab{}.
\newblock \bibinfo{title}{Alya Application}.
\newblock
  \bibinfo{howpublished}{\url{https://www.bsc.es/research-development/research-areas/engineering-simulations/alya-high-performance-computational}}.
\newblock


\bibitem[\protect\citeauthoryear{Haj-Yahya, Alser, Kim, Yağlıkçı,
  Vijaykumar, Rotem, and Mutlu}{Haj-Yahya et~al\mbox{.}}{2020}]%
        {jawad}
\bibfield{author}{\bibinfo{person}{Jawad Haj-Yahya}, \bibinfo{person}{Mohammed
  Alser}, \bibinfo{person}{Jeremie Kim}, \bibinfo{person}{A.~Giray
  Yağlıkçı}, \bibinfo{person}{Nandita Vijaykumar}, \bibinfo{person}{Efraim
  Rotem}, {and} \bibinfo{person}{Onur Mutlu}.} \bibinfo{year}{2020}\natexlab{}.
\newblock \showarticletitle{SysScale: Exploiting Multi-domain Dynamic Voltage
  and Frequency Scaling for Energy Efficient Mobile Processors}. In
  \bibinfo{booktitle}{\emph{2020 ACM/IEEE 47th Annual International Symposium
  on Computer Architecture (ISCA)}}.
\newblock


\bibitem[\protect\citeauthoryear{Han, Park, and Baek}{Han
  et~al\mbox{.}}{2021}]%
        {HMP/DVFS/Criticality/Ratio/Prediction/TPDS2020}
\bibfield{author}{\bibinfo{person}{Myeonggyun Han}, \bibinfo{person}{Jinsu
  Park}, {and} \bibinfo{person}{Woongki Baek}.}
  \bibinfo{year}{2021}\natexlab{}.
\newblock \showarticletitle{Design and Implementation of a Criticality- and
  Heterogeneity-Aware Runtime System for Task-Parallel Applications}.
\newblock \bibinfo{journal}{\emph{IEEE TPDS}} (\bibinfo{year}{2021}).
\newblock


\bibitem[\protect\citeauthoryear{Herbert and Marculescu}{Herbert and
  Marculescu}{2007}]%
        {PerCorevsClustered}
\bibfield{author}{\bibinfo{person}{Sebastian Herbert} {and}
  \bibinfo{person}{Diana Marculescu}.} \bibinfo{year}{2007}\natexlab{}.
\newblock \showarticletitle{Analysis of dynamic voltage/frequency scaling in
  chip-multiprocessors}. In \bibinfo{booktitle}{\emph{Proceedings of the 2007
  international symposium on Low power electronics and design (ISLPED '07)}}.
\newblock


\bibitem[\protect\citeauthoryear{Holmbacka and Keller}{Holmbacka and
  Keller}{2017}]%
        {TaskAwarenessSchedule/ICA3PP2017}
\bibfield{author}{\bibinfo{person}{Simon Holmbacka} {and}
  \bibinfo{person}{J{\"o}rg Keller}.} \bibinfo{year}{2017}\natexlab{}.
\newblock \showarticletitle{Workload Type-Aware Scheduling on big.LITTLE
  Platforms}. In \bibinfo{booktitle}{\emph{Algorithms and Architectures for
  Parallel Processing}}.
\newblock


\bibitem[\protect\citeauthoryear{Isci, Contreras, and Martonosi}{Isci
  et~al\mbox{.}}{2006}]%
        {Phases/MPKI/offline/DVFS/Micro2006}
\bibfield{author}{\bibinfo{person}{Canturk Isci}, \bibinfo{person}{Gilberto
  Contreras}, {and} \bibinfo{person}{Margaret Martonosi}.}
  \bibinfo{year}{2006}\natexlab{}.
\newblock \showarticletitle{Live, Runtime Phase Monitoring and Prediction on
  Real Systems with Application to Dynamic Power Management}. In
  \bibinfo{booktitle}{\emph{2006 39th Annual IEEE/ACM International Symposium
  on Microarchitecture (MICRO'06)}}.
\newblock


\bibitem[\protect\citeauthoryear{Jibaja, Cao, Blackburn, and McKinley}{Jibaja
  et~al\mbox{.}}{2016}]%
        {WASH/regression/PMC/CGO2016}
\bibfield{author}{\bibinfo{person}{Ivan Jibaja}, \bibinfo{person}{Ting Cao},
  \bibinfo{person}{Stephen~M. Blackburn}, {and} \bibinfo{person}{Kathryn~S.
  McKinley}.} \bibinfo{year}{2016}\natexlab{}.
\newblock \showarticletitle{Portable Performance on Asymmetric Multicore
  Processors}. In \bibinfo{booktitle}{\emph{Proceedings of the 2016
  International Symposium on Code Generation and Optimization}}
  \emph{(\bibinfo{series}{CGO '16})}.
\newblock


\bibitem[\protect\citeauthoryear{Moseley, Vachharajani, and Jalby}{Moseley
  et~al\mbox{.}}{2011}]%
        {PMC_Flaws_NPC2011}
\bibfield{author}{\bibinfo{person}{Tipp Moseley}, \bibinfo{person}{Neil
  Vachharajani}, {and} \bibinfo{person}{William Jalby}.}
  \bibinfo{year}{2011}\natexlab{}.
\newblock \showarticletitle{Hardware Performance Monitoring for the Rest of Us:
  A Position and Survey}. In \bibinfo{booktitle}{\emph{Network and Parallel
  Computing}}.
\newblock


\bibitem[\protect\citeauthoryear{Mutlu, Ghose, G{\'o}mez-Luna, and
  Ausavarungnirun}{Mutlu et~al\mbox{.}}{2023}]%
        {MemoryEnergyIncrease}
\bibfield{author}{\bibinfo{person}{Onur Mutlu}, \bibinfo{person}{Saugata
  Ghose}, \bibinfo{person}{Juan G{\'o}mez-Luna}, {and} \bibinfo{person}{Rachata
  Ausavarungnirun}.} \bibinfo{year}{2023}\natexlab{}.
\newblock \showarticletitle{A modern primer on processing in memory}.
\newblock In \bibinfo{booktitle}{\emph{Emerging Computing: From Devices to
  Systems}}. \bibinfo{publisher}{Springer}, \bibinfo{pages}{171--243}.
\newblock


\bibitem[\protect\citeauthoryear{Navarro Mu\~{n}oz, F.~Lorenzon,
  Ayguad\'{e}~Parra, and Beltran~Querol}{Navarro Mu\~{n}oz
  et~al\mbox{.}}{2021}]%
        {DCT/DVFS/Task-base/ICPP21}
\bibfield{author}{\bibinfo{person}{Antoni Navarro Mu\~{n}oz},
  \bibinfo{person}{Arthur F.~Lorenzon}, \bibinfo{person}{Eduard
  Ayguad\'{e}~Parra}, {and} \bibinfo{person}{Vicen\c{c} Beltran~Querol}.}
  \bibinfo{year}{2021}\natexlab{}.
\newblock \showarticletitle{Combining Dynamic Concurrency Throttling with
  Voltage and Frequency Scaling on Task-Based Programming Models}. In
  \bibinfo{booktitle}{\emph{50th International Conference on Parallel
  Processing}} \emph{(\bibinfo{series}{ICPP 2021})}.
\newblock


\bibitem[\protect\citeauthoryear{{OpenMP Architecture Review Board}}{{OpenMP
  Architecture Review Board}}{2018}]%
        {openmp50-api}
\bibfield{author}{\bibinfo{person}{{OpenMP Architecture Review Board}}.}
  \bibinfo{year}{2018}\natexlab{}.
\newblock \bibinfo{title}{OpenMP Application Program Interface. Version 5.0}.
\newblock
\newblock


\bibitem[\protect\citeauthoryear{Patki, Lowenthal, Rountree, Schulz, and
  de~Supinski}{Patki et~al\mbox{.}}{2013}]%
        {MaxPerf/Powerlimit/ICS13}
\bibfield{author}{\bibinfo{person}{Tapasya Patki}, \bibinfo{person}{David~K.
  Lowenthal}, \bibinfo{person}{Barry Rountree}, \bibinfo{person}{Martin
  Schulz}, {and} \bibinfo{person}{Bronis~R. de Supinski}.}
  \bibinfo{year}{2013}\natexlab{}.
\newblock \showarticletitle{Exploring Hardware Overprovisioning in
  Power-Constrained, High Performance Computing}. In
  \bibinfo{booktitle}{\emph{Proceedings of the 27th International ACM
  Conference on International Conference on Supercomputing}}
  \emph{(\bibinfo{series}{ICS '13})}. \bibinfo{publisher}{Association for
  Computing Machinery}, \bibinfo{address}{New York, NY, USA},
  \bibinfo{pages}{173–182}.
\newblock
\showISBNx{9781450321303}


\bibitem[\protect\citeauthoryear{Rauber and Rünger}{Rauber and
  Rünger}{2018}]%
        {SchedDec/DVFS/CPE2019}
\bibfield{author}{\bibinfo{person}{Thomas Rauber} {and} \bibinfo{person}{Gudula
  Rünger}.} \bibinfo{year}{2018}\natexlab{}.
\newblock \showarticletitle{A scheduling selection process for
  energy‐efficient task execution on DVFS processors}.
\newblock \bibinfo{journal}{\emph{Concurrency and Computation: Practice and
  Experience}}  \bibinfo{volume}{31} (\bibinfo{date}{10} \bibinfo{year}{2018}).
\newblock


\bibitem[\protect\citeauthoryear{Reddy, Singh, Biswas, Merrett, and
  Al-Hashimi}{Reddy et~al\mbox{.}}{2018}]%
        {Asymmetric/Clusterd/MultipleAPP/MRPI}
\bibfield{author}{\bibinfo{person}{Basireddy~Karunakar Reddy},
  \bibinfo{person}{Amit~Kumar Singh}, \bibinfo{person}{Dwaipayan Biswas},
  \bibinfo{person}{Geoff~V. Merrett}, {and} \bibinfo{person}{Bashir~M.
  Al-Hashimi}.} \bibinfo{year}{2018}\natexlab{}.
\newblock \showarticletitle{Inter-Cluster Thread-to-Core Mapping and DVFS on
  Heterogeneous Multi-Cores}.
\newblock \bibinfo{journal}{\emph{IEEE Transactions on Multi-Scale Computing
  Systems}} \bibinfo{volume}{4}, \bibinfo{number}{3} (\bibinfo{year}{2018}).
\newblock


\bibitem[\protect\citeauthoryear{Ribic and Liu}{Ribic and Liu}{2016}]%
        {HarisRibic-AEQUITAS}
\bibfield{author}{\bibinfo{person}{Haris Ribic} {and} \bibinfo{person}{Yu
  Liu}.} \bibinfo{year}{2016}\natexlab{}.
\newblock \showarticletitle{AEQUITAS: Coordinated Energy Management Across
  Parallel Applications}. In \bibinfo{booktitle}{\emph{2016 ACM International
  Conference on Supercomputing}}. \bibinfo{pages}{1--12}.
\newblock


\bibitem[\protect\citeauthoryear{Ribic and Liu}{Ribic and Liu}{2014}]%
        {HarisRibic-WorkStealing/PercoreDVFS/Tempo}
\bibfield{author}{\bibinfo{person}{Haris Ribic} {and} \bibinfo{person}{Yu~David
  Liu}.} \bibinfo{year}{2014}\natexlab{}.
\newblock \showarticletitle{Energy-Efficient Work-Stealing Language Runtimes}.
  In \bibinfo{booktitle}{\emph{Proceedings of the 19th International Conference
  on Architectural Support for Programming Languages and Operating Systems}}
  \emph{(\bibinfo{series}{ASPLOS ’14})}.
\newblock


\bibitem[\protect\citeauthoryear{Rotem, Mandelblat, Basin, Weissmann, Gihon,
  Chabukswar, Fenger, and Gupta}{Rotem et~al\mbox{.}}{2021}]%
        {IntelAlderLake}
\bibfield{author}{\bibinfo{person}{Efraim Rotem}, \bibinfo{person}{Yuli
  Mandelblat}, \bibinfo{person}{Vadim Basin}, \bibinfo{person}{Eli Weissmann},
  \bibinfo{person}{Arik Gihon}, \bibinfo{person}{Rajshree Chabukswar},
  \bibinfo{person}{Russ Fenger}, {and} \bibinfo{person}{Monica Gupta}.}
  \bibinfo{year}{2021}\natexlab{}.
\newblock \showarticletitle{Alder Lake Architecture}. In
  \bibinfo{booktitle}{\emph{2021 IEEE Hot Chips 33 Symposium (HCS)}}.
\newblock


\bibitem[\protect\citeauthoryear{Sagi, Doan, Rapp, Wild, Henkel, and
  Herkersdorf}{Sagi et~al\mbox{.}}{2020}]%
        {Regression/powermodel/TCAD2020}
\bibfield{author}{\bibinfo{person}{Mark Sagi}, \bibinfo{person}{Nguyen Anh~Vu
  Doan}, \bibinfo{person}{Martin Rapp}, \bibinfo{person}{Thomas Wild},
  \bibinfo{person}{Jörg Henkel}, {and} \bibinfo{person}{Andreas Herkersdorf}.}
  \bibinfo{year}{2020}\natexlab{}.
\newblock \showarticletitle{A Lightweight Nonlinear Methodology to Accurately
  Model Multicore Processor Power}.
\newblock \bibinfo{journal}{\emph{IEEE Transactions on Computer-Aided Design of
  Integrated Circuits and Systems}} \bibinfo{volume}{39}, \bibinfo{number}{11}
  (\bibinfo{year}{2020}).
\newblock


\bibitem[\protect\citeauthoryear{Shafik, Das, Yang, Merrett, and
  Al-Hashimi}{Shafik et~al\mbox{.}}{2015}]%
        {EnergyMinimization/OpenMP/DVFS/DCT/PARMA-DITAM2015}
\bibfield{author}{\bibinfo{person}{Rishad~A. Shafik}, \bibinfo{person}{Anup
  Das}, \bibinfo{person}{Sheng Yang}, \bibinfo{person}{Geoff Merrett}, {and}
  \bibinfo{person}{Bashir~M. Al-Hashimi}.} \bibinfo{year}{2015}\natexlab{}.
\newblock \showarticletitle{Adaptive Energy Minimization of OpenMP Parallel
  Applications on Many-Core Systems}. In \bibinfo{booktitle}{\emph{Proceedings
  of the 6th Workshop on Parallel Programming and Run-Time Management
  Techniques for Many-Core Architectures}} \emph{(\bibinfo{series}{PARMA-DITAM
  '15})}.
\newblock


\bibitem[\protect\citeauthoryear{Simonyan and Zisserman}{Simonyan and
  Zisserman}{2014}]%
        {VGG16}
\bibfield{author}{\bibinfo{person}{Karen Simonyan} {and}
  \bibinfo{person}{Andrew Zisserman}.} \bibinfo{year}{2014}\natexlab{}.
\newblock \showarticletitle{Very deep convolutional networks for large-scale
  image recognition}.
\newblock \bibinfo{journal}{\emph{arXiv preprint arXiv:1409.1556}}
  (\bibinfo{year}{2014}).
\newblock


\bibitem[\protect\citeauthoryear{Sundriyal and Sosonkina}{Sundriyal and
  Sosonkina}{2016}]%
        {JointDVFS/Supercomputing16}
\bibfield{author}{\bibinfo{person}{Vaibhav Sundriyal} {and}
  \bibinfo{person}{Masha Sosonkina}.} \bibinfo{year}{2016}\natexlab{}.
\newblock \showarticletitle{Joint Frequency Scaling of Processor and DRAM}.
\newblock \bibinfo{journal}{\emph{J. Supercomput.}} \bibinfo{volume}{72},
  \bibinfo{number}{4} (\bibinfo{date}{apr} \bibinfo{year}{2016}),
  \bibinfo{pages}{1549–1569}.
\newblock
\showISSN{0920-8542}


\bibitem[\protect\citeauthoryear{Torng, Wang, and Batten}{Torng
  et~al\mbox{.}}{2016}]%
        {Christopher-ISCA2016-Asymmetric/PercoreDVFS/Parallelism}
\bibfield{author}{\bibinfo{person}{Christopher Torng}, \bibinfo{person}{Moyang
  Wang}, {and} \bibinfo{person}{Christopher Batten}.}
  \bibinfo{year}{2016}\natexlab{}.
\newblock \showarticletitle{Asymmetry-Aware Work-Stealing Runtimes}. In
  \bibinfo{booktitle}{\emph{2016 ACM/IEEE 43rd Annual International Symposium
  on Computer Architecture (ISCA)}}. \bibinfo{pages}{40--52}.
\newblock


\bibitem[\protect\citeauthoryear{Wu, Taylor, Cook, and Mucci}{Wu
  et~al\mbox{.}}{2016}]%
        {Regression/PMC/Perf/Power/MC2016}
\bibfield{author}{\bibinfo{person}{Xingfu Wu}, \bibinfo{person}{Valerie
  Taylor}, \bibinfo{person}{Jeanine Cook}, {and} \bibinfo{person}{Philip~J.
  Mucci}.} \bibinfo{year}{2016}\natexlab{}.
\newblock \showarticletitle{Using Performance-Power Modeling to Improve Energy
  Efficiency of HPC Applications}.
\newblock \bibinfo{journal}{\emph{Computer}} \bibinfo{volume}{49},
  \bibinfo{number}{10} (\bibinfo{year}{2016}), \bibinfo{pages}{20--29}.
\newblock


\end{thebibliography}
\end{document}